\documentclass[12pt,letterpaper]{article}
\usepackage[usenames, dvipsnames]{xcolor}
\usepackage{jcapmod}

\usepackage{tocloft}
\usepackage[]{todonotes}
\usepackage{verbatim}
\usepackage{mathrsfs}
\usepackage{cleveref}
\usepackage[shortlabels]{enumitem}
\usepackage{pgf}
\usepackage{tikz-cd}
\usetikzlibrary{shapes,arrows}
\usetikzlibrary{calc}
\usepackage{pgfplots}
\pgfplotsset{compat=1.12}
\usepackage{tikz-3dplot}

\usepackage{tikz}
\usepackage{setspace,caption}
\usepackage{lipsum}
\usetikzlibrary{matrix,arrows,calc}
\makeatletter
\newsavebox\myboxA
\newsavebox\myboxB
\newlength\mylenA

\definecolor{cornellRed}{HTML}{B31B1B}
\definecolor{cornellBlue}{HTML}{0068AC}
\definecolor{cornellGreen}{HTML}{6EB43F}

\usepackage{bbm} 					
\usepackage{slashed} 				
\usepackage{graphicx}				
\usepackage{subcaption}			
\usepackage{psfrag}				
\usepackage{tensor}				
\usepackage{fouridx}				
\usepackage{bm}					
\usepackage{mdframed}				
\usepackage{multirow}				
\usepackage{soul}					
\usepackage{bbold}				
\usepackage{multicol}				
\usepackage{tikz-cd}
\usepackage{rotating}

\usetikzlibrary{arrows}

\tikzset{
commutative diagrams/.cd,
arrow style=tikz,
diagrams={>=latex}}

\usepackage{amsmath}
\usepackage{amssymb}
\usepackage{amsfonts}
\usepackage{mathtools}

\usepackage{feynmf}
\usepackage{marvosym}

\usepackage{import}

\newcommand*\xoverline[2][0.75]{%
    \sbox{\myboxA}{$\m@th#2$}%
    \setbox\myboxB\null
    \ht\myboxB=\ht\myboxA%
    \dp\myboxB=\dp\myboxA%
    \wd\myboxB=#1\wd\myboxA
    \sbox\myboxB{$\m@th\overline{\copy\myboxB}$}
    \setlength\mylenA{\the\wd\myboxA}
    \addtolength\mylenA{-\the\wd\myboxB}%
    \ifdim\wd\myboxB<\wd\myboxA%
       \rlap{\hskip 0.5\mylenA\usebox\myboxB}{\usebox\myboxA}%
    \else
        \hskip -0.5\mylenA\rlap{\usebox\myboxA}{\hskip 0.5\mylenA\usebox\myboxB}%
    \fi}
\makeatother

\newcommand{\overbar}[1]{\mkern 1.5mu\overline{\mkern-1.5mu#1\mkern-1.5mu}\mkern 1.5mu}

\newcommand{\diff}{\partial}

\newcommand{\im}{\,\mathrm{Im}\,}
\newcommand{\re}{\,\mathrm{Re}\,}

\newcommand{\tr}{\,\mathrm{tr}}

\definecolor{cobalt}{RGB}{44, 98, 120}
\definecolor{celadon}{rgb}{0.67, 0.88, 0.69}
\definecolor{dm}{cmyk}{.20, 0, .30, 0}
\definecolor{burgundy}{rgb}{0.5, 0.0, 0.13}
\definecolor{plotBlue}{RGB}{94, 130, 181}

\hypersetup{
  colorlinks,
  citecolor=Violet,
  linkcolor=cobalt,
  urlcolor=Blue}

\DeclareSymbolFontAlphabet{\mathbb}{AMSb}

\newif\iffastcompile

\fastcompilefalse

\iffastcompile
\newcommand{\mk}[1]{}
\newcommand{\lm}[1]{}
\else
\newcommand{\mk}[1]{\todo[color=burgundy!30, size=\scriptsize, bordercolor=burgundy!30]{MK: #1}}
\newcommand{\lm}[1]{\todo[color=dm!90, size=\scriptsize, bordercolor=dm!90]{LM: #1}}

\fi

\ProvideTextCommandDefault{\Dbar}{%
\leavevmode\lower.5ex\rlap{\hskip-.07em\accent"16}D%
}

\begin{document}
	\newcommand{\main}{.}
\begin{titlepage}

\setcounter{page}{1} \baselineskip=15.5pt \thispagestyle{empty}
\setcounter{tocdepth}{1}

\bigskip\

\vspace{1cm}
\begin{center}
{\fontsize{22}{28} \bfseries
de Sitter Vacua  from Ten Dimensions}
\end{center}

\vspace{0.55cm}

\begin{center}
\scalebox{0.95}[0.95]{{\fontsize{14}{30}\selectfont Shamit Kachru,$^{a}$ Manki Kim,$^{b}$ Liam McAllister,$^{b}$ and Max Zimet$^{a}$\vspace{0.25cm}}}

\end{center}

\begin{center}

\vspace{0.15 cm}
{\fontsize{11}{30}
\noindent\textsl{$^{a}$Stanford Institute for Theoretical Physics, Stanford University, Stanford, CA 94305}\\
\textsl{$^{b}$Department of Physics, Cornell University, Ithaca, NY 14853}}\\

\vspace{0.25cm}

\vskip .5cm
\end{center}

\begin{center}
------ \emph{Dedicated to the memory of Steven S.~Gubser} ------
\end{center}

\vspace{0.8cm}
\noindent

We analyze the de Sitter construction of \cite{KKLT} using ten-dimensional supergravity, finding exact agreement with the four-dimensional effective theory.
Starting from the fermionic couplings in the D7-brane action, we derive the ten-dimensional stress-energy due to gaugino condensation on D7-branes.
We demonstrate that upon including this stress-energy, as well as that due to anti-D3-branes, the ten-dimensional equations of motion require the four-dimensional curvature to
take precisely the value determined by the four-dimensional effective theory of \cite{KKLT}.

\vspace{1.1cm}

\vspace{3.1cm}

\noindent\today

\end{titlepage}
\tableofcontents\newpage

\section{Introduction}

A foundational problem in cosmology is to characterize de Sitter solutions of string theory.  Tremendous efforts have been expended in the study of flux compactifications of weakly-coupled type II string theories on orientifolds (see e.g.~the reviews \cite{Silverstein:2004id,Grana:2005jc,sk:fluxes,Blumenhagen:2006ci,sk:fluxes2,McAllister:2007bg,Denef:2008wq,Baumann:2014nda,quevedo:review}).
Non-supersymmetric vacua necessarily remain more difficult to analyze than supersymmetric ones, if only because fewer theoretical tools can be applied there.
However, we can take heart by recalling that the entirety of real-world physics is strictly non-supersymmetric, and progress has nonetheless been possible in a few areas, beginning with the work of the non-supersymmetric theorists of antiquity.

A paradigm for exhibiting realistic compactifications of string theory is to derive directly the properties of a four-dimensional effective theory in parametrically controlled limits, such as weak coupling, large volume, and small supersymmetry breaking, and then carefully argue for the form of corrections to the effective theory away from such limits.  When the corrections are parametrically small, one expects the vacuum structure computed in the effective theory to be robust.

The couplings in such an effective theory can sometimes be computed in more than one way, e.g.~on the string worldsheet and in ten-dimensional supergravity.  When dual perspectives are available, they provide a cross-check that lends a degree of further support to the computation of the effective theory.  However, it is rarely the case that everything that can be computed in one duality frame can also be computed in the other frame: instead, certain effects are manifest in one frame, and other effects are manifest in the other frame, as is familiar from famous strong-weak dualities in quantum field theory and holography.

The study of de Sitter vacua of type IIB string theory compactified on orientifolds of Calabi-Yau threefolds, as in \cite{KKLT}, has relied heavily on computations of vacuum structure in the four-dimensional effective theory.  However, certain questions about these theories are intrinsically ten-dimensional, and answering them requires a quantitative description of the de Sitter vacua in terms of configurations of ten-dimensional fields.  For example, integrating the ten-dimensional equations of motion over the compact space reveals constraints on possible solutions (see e.g.~\cite{DeWit,Giddings:2001yu,Steinhardt:2008nk,Moritz:2017xto}), and it would be instructive to expose all such constraints.  Similarly, the couplings between distinct sectors of the effective theory are often most readily computed by finding solutions for the massless fields in ten dimensions.

At the same time, it is not generally possible even in principle to derive all four-dimensional couplings through a purely ten-dimensional computation.  Consider, for example, the infrared dynamics of a pure $\mathcal{N}=1$ super-Yang-Mills theory arising on a collection of D7-branes that wrap a four-cycle $\Sigma$ in the compact space.  The eight-dimensional gauge theory is not even asymptotically free, but at energies far below the Kaluza-Klein scale, the four-dimensional theory confines and generates a gaugino condensate.  Attempting to compute the gaugino condensate from the ten-dimensional equations of motion, and rejecting the simplifications of the four-dimensional description, would be quixotically self-limiting.

A practical approach, then, is to compute the configuration of ten-dimensional fields that corresponds to a four-dimensional de Sitter vacuum, while taking specific expectation values --- such as those of gaugino bilinears --- to be those determined by the four-dimensional equations of motion.
We refer to the result of this analysis as a \emph{ten-dimensional description} of a de Sitter vacuum.

In this work we provide a ten-dimensional description of the de Sitter scenario of \cite{KKLT}.
This problem has been examined in \cite{Moritz:2017xto,Hamada:2018qef,Gautason:2018gln,Carta:2019rhx,Gautason:2019jwq,Hamada:2019ack,Bena:2019mte} (see also the earlier works \cite{Koerber:2007xk,Baumann:2010sx,Dymarsky:2010mf}).
As we will explain below, our analysis aligns with some aspects of these works, but also resolves certain puzzles that were implicit in the literature.

Our approach is a computation from an elementary starting point.
Beginning with the ten-dimensional action of type I string theory, we derive the two-gaugino and four-gaugino couplings on D7-branes, and then compute the ten-dimensional stress-energy sourced by a gaugino bilinear expectation value $\langle\lambda\lambda\rangle$.
Then, taking $\langle\lambda\lambda\rangle$ to have the value predicted by the four-dimensional super-Yang-Mills theory --- and we stress that this step is the only point at which information from four dimensions is injected --- we compute the four-dimensional scalar curvature determined by the ten-dimensional equations of motion.

In order to evaluate the contribution of the D7-brane gaugino-flux coupling, we use the Killing spinor equations for compactification on a generalized complex geometry, with which we establish that in a supersymmetric configuration the generalized complex geometry superpotential equals the full superpotential of the four-dimensional theory.
We then compare the scalar curvature resulting from the ten-dimensional configuration to the scalar curvature determined by the four-dimensional Einstein equations equipped with the scalar potential of \cite{KKLT}.
We prove that the match is exact in the supersymmetric vacuum.  Furthermore, provided that
the generalized complex geometry superpotential continues to equal the full superpotential in off-shell configurations --- which we find very plausible but do not prove here --- our ten-dimensional computation of the scalar potential for the K\"ahler modulus continues to precisely match the four-dimensional theory, in the presence of anti-D3-branes as well as off-shell.

The organization of this paper is as follows.
In \S\ref{sec:EOM} we assemble the equations of motion of type IIB supergravity.
In \S\ref{sec:SUSY} we consider the effects of an expectation value for the gaugino bilinear on a stack of D7-branes.
We show that couplings of the D7-brane gauginos, including the couplings to flux derived by Dymarsky and Martucci in \cite{Dymarsky:2010mf} following \cite{Camara:2004jj},
source a contribution $T_{\mu\nu}^{\langle\lambda\lambda\rangle}$ to the stress-energy tensor.  Including
this stress-energy in the ten-dimensional equations of motion, we compute the four-dimensional scalar curvature, and find perfect agreement with that determined by
the F-term potential in the four-dimensional $\mathcal{N}=1$ supersymmetric effective theory of
\cite{KKLT}.
In \S\ref{sec:dS} we consider the combined effects of an anti-D3-brane and a D7-brane gaugino bilinear.
We examine the ten-dimensional supergravity solution with these sources and show that $T_{\mu\nu}^{\langle\lambda\lambda\rangle}$ continues to match the four-dimensional potential derived in \cite{KKLT}.
Our conclusions appear in \S\ref{sec:conclusions}.
In Appendix \ref{app:conv} we first dimensionally reduce and T-dualize the type I action to obtain the couplings of D7-brane gauginos.
We then analyze the ten-dimensional Killing spinor equations, correcting an inconsistency in the literature, and use them to demonstrate explicitly that the superpotential for compactification on a generalized complex geometry captures both the classical flux superpotential and the gaugino condensate superpotential.
Appendix \ref{app:spec} shows, based on the spectroscopy of $T^{1,1}$, that the interactions of an anti-D3-brane and a gaugino condensate mediated by Kaluza-Klein excitations of a Klebanov-Strassler throat can be neglected compared to the
interaction mediated by the K\"ahler modulus.
In Appendix \ref{app:d3} we consider the singular contributions to the four-dimensional equations of motion, which originate in the fact that the D7-brane stack is localized to a divisor.  We show that these divergent terms cancel each other, and the finite remainder is the four-dimensional scalar potential.  We then repeat this computation for a compactification containing a D3-brane, with analogous results.

\section{Ten-dimensional Equations of Motion} \label{sec:EOM}

In this section, we set our notation and collect useful forms of the ten-dimensional Einstein equations and five-form Bianchi identity.
We then express the stress-energy tensor of the four-dimensional effective theory in terms of the ten-dimensional field configuration.

We consider type IIB string theory on $X\times M,$ where $X$ is a four-dimensional spacetime and $M$ is a six-dimensional compact manifold that in the leading approximation is an O3/O7 orientifold of a Calabi-Yau threefold.  We take the metric ansatz
\begin{equation}
ds^2=G_{AB}dX^AdX^B=e^{-6u(x)+2A(y)}g_{\mu\nu}dx^\mu dx^\nu+e^{2u(x)-2A(y)}g_{ab}dy^a dy^b\,,\label{eqn:metric ansatz}
\end{equation}
with $x$ denoting coordinates in $X$ and $y$ denoting coordinates in $M.$  Greek indices take values in $\{0,\ldots,3\}$, and Latin indices take values in $\{1,\ldots,6\}$.
We use the abbreviations $g_6=\det g_{ab}$ and $g_4=\det g_{\mu\nu}$, and note that $\sqrt{-G}=\sqrt{-g}e^{-6u-2A}=\sqrt{-g_4 g_6}e^{-6u -2A}$.

The ten-dimensional type IIB supergravity action is
\begin{equation}\label{sfullis}
S=\frac{1}{2\kappa_{10}^2}\int \mathrm{d}^{10} X \sqrt{-G}\Biggl(\mathcal{R}_{10}-\frac{\partial_A \tau \partial^A\overbar{\tau}}{2\,(\mathrm{Im}\,\tau)^2}- \frac{G_3\cdot\overbar{G}_3}{2\,\mathrm{Im}\,\tau} -\frac{\tilde{F}_5^2}{4}\Biggr)+\frac{1}{8i\kappa_{10}^2}\int \frac{C_4\wedge G_3\wedge \overbar{G}_3}{\mathrm{Im}\,\tau} + S_{\mathrm{local}}\,,
\end{equation} where $\mathcal{R}_{10}$ is the Ricci scalar computed from $G$, $\tau = C_0 + i\,e^{-\phi}$ is the axiodilaton, $G_3 := F_3 -\tau H_3 \equiv \mathrm{d}C_2-\tau \mathrm{d}B_2$, and $\tilde{F}_5=F_5-\frac{1}{2} C_2 \wedge H_3 + \frac{1}{2} B_2 \wedge F_3$, with $F_5=\mathrm{d}C_4$.  The local term $S_{\mathrm{local}}$ encodes the contributions of D-branes and orientifold planes.
We work in units where $(2\pi)^2\alpha'=1$.

For the five-form $\tilde{F}_5$ we take the ansatz
\begin{equation}
\tilde{F}_5 = (1+\star_{10})e^{-12u}\sqrt{-g_4}\,\mathrm{d}\alpha(y)\wedge\mathrm{d}x^0\wedge\mathrm{d}x^1\wedge\mathrm{d}x^2\wedge\mathrm{d}x^3\,,
\end{equation} with $\star_{10}$ the ten-dimensional Hodge star, and define the scalars
\begin{equation}
\Phi_{\pm}:=e^{4A}\pm\alpha\,.
\end{equation}
We also define the imaginary self-dual and imaginary anti-self-dual fluxes
\begin{equation}
G_{\pm}:=\frac{(\star_6\pm i)}{2}G_3\,,
\end{equation} with $\star_{6}$ the six-dimensional Hodge star.
We abbreviate \eqref{sfullis} as
\begin{equation}\label{sis}
S=\frac{1}{2\kappa_{10}^2}\int\mathrm{d}^{10}X \sqrt{-G}\,\mathcal{R}_{10}+\int \mathrm{d}^{10}X\mathcal{L} \equiv S_{\mathrm{EH}}+\int \mathrm{d}^{10}X\mathcal{L}\,,
\end{equation}
with $\mathcal{L}$ encoding everything except for the Einstein-Hilbert term.

From \eqref{eqn:metric ansatz} one computes the Ricci tensors
\begin{equation}
\mathcal{R}_{4,\mu\nu}=\mathcal{R}_{4,\mu\nu}[g]-e^{-8u+4A}g_{\mu\nu}\nabla^2A+3g_{\mu\nu}\,\square u -24\diff_\mu u\diff_\nu u,\label{eqn:external Ricci}
\end{equation}
\begin{equation}
\mathcal{R}_{6,ab}=\mathcal{R}_{6,ab}[g]+\nabla^2 Ag_{ab}-e^{8u-4A}\,g_{ab}\,\square u -8\diff_a A\diff_b A\,,\label{eqn:internal Ricci}
\end{equation}
where $\mathcal{R}_{4,\mu\nu}[g]$ and $\mathcal{R}_{6,ab}[g]$ are the Ricci tensors of $g_{\mu\nu}$ and $g_{ab}$, respectively.
Expanding the Einstein-Hilbert part of \eqref{sfullis} using \eqref{eqn:external Ricci} and \eqref{eqn:internal Ricci}, we find
\begin{equation}
S_{\mathrm{EH}}=\frac{1}{2\kappa_{10}^2}\int d^4x\, d^6y \sqrt{-g_4 g_6} \Bigl(e^{-4A}\mathcal{R}_4[g]+e^{-8u}\mathcal{R}_6[g]-24e^{-4A}\diff_\mu u\diff^\mu u-8e^{-8u}\diff_a A\diff^a A\Bigr)\,, \notag
\end{equation} where indices are raised using $g_{\mu\nu}$ or $g_{ab}$ as appropriate.
The Planck mass is given by
\begin{equation}
M_{\mathrm{pl}}^2=\frac{\mathcal{V}}{\kappa_{10}^2}\,,
\end{equation}
where $\mathcal{V}$ is the warped volume of $M$, defined as
\begin{equation}
\mathcal{V}=\int_Md^6y\sqrt{g_6}e^{-4A}\,.
\end{equation}

The equation of motion for the breathing mode $u$ obtained from \eqref{sis} is
\begin{equation}
24\, \square u=4 e^{4A-8u}\Bigl(\mathcal{R}_6[g]-8\diff_a A\diff^a A\Bigr)-\kappa_{10}^2 e^{4A}\frac{\delta \mathcal{L}}{\delta u}\,.\label{eqn:eom u 1}
\end{equation}
We next turn to the Einstein equations, in conventions where the stress-energy tensor is defined as
\begin{equation}
T_{AB}=-\frac{2}{\sqrt{-G}}\frac{\delta\mathcal{L}}{\delta G^{AB}}\,.
\end{equation}
The four-dimensional components of the ten-dimensional Einstein equations are
\begin{equation}
\mathcal{R}_{4,\mu\nu}=\kappa_{10}^2\Bigl( T_{\mu\nu}-\frac{1}{8}G_{\mu\nu} T\Bigr)\,.\label{eqn:Einstein 1}
\end{equation}
Reversing the trace using the ten-dimensional metric $G^{\mu\nu}$, we have
\begin{equation}
\mathcal{R}_{4,\mu\nu}G^{\mu\nu}=-\kappa_{10}^2 T_{\mu\nu}G^{\mu\nu}-\frac{\kappa_{10}^2}{2}\Bigl(T_{ab}G^{ab}-3T_{\mu\nu}G^{\mu\nu}\Bigr)\,.\label{eqn:Einstein 2}
\end{equation}
Integrating \eqref{eqn:Einstein 2} over $M$ and using \eqref{eqn:external Ricci} leads to
\begin{equation}
M_{\mathrm{pl}}^2\Bigl(\mathcal{R}_4[g]+12\,\square u-24\diff_\mu u \diff^\mu u\Bigr)=\int_M \sqrt{g_6}e^{-6u-2A}\Bigl[-T_{\mu\nu}G^{\mu\nu}-\frac{1}{2}\left(T_{ab}G^{ab}-3T_{\mu\nu}G^{\mu\nu}\right) \Bigr]\,.\label{eqn:The Einstein}
\end{equation}
Similarly, the six-dimensional components of the ten-dimensional Einstein equations are
\begin{equation}
\mathcal{R}_{6,ab}=\kappa_{10}^2\Bigl(T_{ab}-\frac{1}{8}G_{ab}T\Bigr)\,,
\end{equation}
with trace-reversed form
\begin{equation}
\mathcal{R}_{6,ab}G^{ab}=\frac{\kappa_{10}^2}{4}\Bigl(T_{ab}G^{ab}-3T_{\mu\nu}G^{\mu\nu} \Bigr)\,.\label{eqn:Einstein 3}
\end{equation}
Integrating \eqref{eqn:Einstein 3} over $M$ and using \eqref{eqn:internal Ricci} gives
\begin{equation}
-6M_{\mathrm{pl}}^2\,\square u +\frac{1}{\kappa_{10}^2}\int_{M}\sqrt{g_6} e^{-8u}\Bigl(\mathcal{R}_6[g]-8\diff_aA\diff^a A\Bigr)=\frac{1}{4}\int_M \sqrt{g_6} e^{-6u-2A}\left[T_{ab}G^{ab}-3T_{\mu\nu}G^{\mu\nu} \right]\,.\label{eqn:Einstein 4}
\end{equation}
Finally, we examine the Bianchi identity
\begin{equation}
d\tilde{F}_5=2\mu_3\kappa_{10}^2\rho_{D3}\,\text{dVol}_M = \,H\wedge F+ 2\mu_3\kappa_{10}^2\rho_{D3}^{\mathrm{loc}}\,\text{dVol}_M\,.\label{eqn:Bianchi 1}
\end{equation}
Here $\text{dVol}_M=\sqrt{g_6}dy^1\wedge \cdots \wedge dy^6$,
$\rho_{D3}$ is the net
D3-brane charge density, and $\rho_{D3}^{\mathrm{loc}}$ is the net D3-brane charge density of localized objects such as D3-branes and anti-D3-branes.  (We use $\rho_{\overbar{D3}}$ to denote the contributions of anti-D3-branes specifically.)
From \eqref{eqn:Bianchi 1} we derive  the useful integrated form
\begin{align}
0=\int_M \sqrt{g_6} \left(e^{-8u-8A}\diff_a e^{4A}\diff^a \alpha+2\mu_3\kappa_{10}^2e^{-12u}e^{4A}\rho_{D3}\right).\label{eqn:Bianchi 2}
\end{align}
Combining \eqref{eqn:The Einstein},  \eqref{eqn:Einstein 4}, and \eqref{eqn:Bianchi 2}
we obtain
\begin{equation}
\begin{aligned}\label{eqn:Master 2}
M_{\mathrm{pl}}^2\mathcal{R}_4[g] &= 24M_{\mathrm{pl}}^2\diff_\mu u\diff^\mu u-\int_M\sqrt{g_6} \Bigl(e^{-4A}\hat{T}_{\mu\nu}g^{\mu\nu}+4\mu_3e^{-12u+4A}\rho_{D3}\Bigr)\\ &-\frac{2e^{-8u}}{\kappa_{10}^2}\int_M\sqrt{g_6}\, \mathcal{R}_6[g] +\frac{e^{-8u}}{\kappa_{10}^2}\int_M\sqrt{g_6}e^{-8A}\diff_a\Phi_-\diff^a\Phi_-\,,
\end{aligned}
\end{equation}
where $\hat{T}_{\mu\nu}$ denotes the  stress-energy tensor  excluding  the contribution from $\tilde{F}_5.$

Substituting the type IIB supergravity action \eqref{sfullis} into \eqref{eqn:Master 2}, and taking $S_{\mathrm{local}}$ in \eqref{sfullis} to include D3-branes and D7-branes,
we find
\begin{equation}
\begin{aligned}\label{eqn:Master Final}
M_{\mathrm{pl}}^2\mathcal{R}_4[g]&=24M_{\mathrm{pl}}^2\diff_\mu u\diff^\mu u+\frac{\diff_\mu\tau\diff^\mu\overbar{\tau}}{(\im\tau)^2}+8\mu_3\int_M \sqrt{g_6} e^{-12u+4A}\rho_{\overbar{D3}}-\int_M \sqrt{g_6} e^{-4A}T_{\mu\nu}^{D7}g^{\mu\nu} \\
&-\frac{2e^{-8u}}{\kappa_{10}^2}\int_M \sqrt{g_6}\mathcal{R}_6[g]+\frac{e^{-8u}}{\kappa_{10}^2}\int_M \sqrt{g_6}e^{-8A}\diff_a \Phi_-\diff^a\Phi_-\,.
\end{aligned}
\end{equation}
To interpret \eqref{eqn:Master Final}, we consider a general four-dimensional action
\begin{equation}
S_{4}=\frac{M_{\mathrm{pl}}^2}{2}\int_{X}\sqrt{-g_4}\,\mathcal{R}_4[g]+\int_X \sqrt{-g_4}\, \mathcal{L}_4\,.
\end{equation}
The four-dimensional Einstein equations imply
\begin{equation}\label{ee4}
M_{\mathrm{pl}}^2 \mathcal{R}_{4}[g]=-\mathscr{T}\,,
\end{equation} where $\mathscr{T}_{\mu\nu}$ is the \emph{four-dimensional} stress-energy tensor, i.e.~the stress-energy tensor computed from $\mathcal{L}_4$.
The  four-dimensional stress-energy tensor $\mathscr{T}_{\mu\nu}$ and the four-dimensional components $T_{\mu\nu}$ of the ten-dimensional stress-energy tensor $T_{AB}$ are related by
\begin{align}\label{4d10dt}
\mathscr{T}_{\mu\nu}=&\int_M \sqrt{g_6} \left[ e^{-4A} \hat{T}_{\mu\nu}+\mu_3 e^{4A-12u}g_{\mu\nu}\rho_{D3}+\frac{e^{-8u}}{2\kappa_{10}^2}g_{\mu\nu}\mathcal{R}_6[g]-\frac{e^{-8A-8u}}{4\kappa_{10}^2}g_{\mu\nu} \diff_a\Phi_-\diff^a\Phi_-\right]\nonumber\\
&+M_{\mathrm{pl}}^2\bigl(24\diff_\mu u\diff_\nu u-12g_{\mu\nu}\diff_\rho u\diff^\rho u\bigr).
\end{align}
Comparing \eqref{eqn:Master Final} and \eqref{ee4}, the right-hand side of \eqref{eqn:Master Final} can be identified with $-\mathscr{T}$, i.e.~with minus the trace of the stress-energy tensor of the effective theory.

The master equation \eqref{eqn:Master Final} thus encodes the relationship between the curvature $\mathcal{R}_4[g]$ of the four-dimensional Einstein frame metric $g_{\mu\nu}$ on the one hand, and the contributions of the ten-dimensional field configuration to the effective four-dimensional stress-energy tensor $\mathscr{T}_{\mu\nu}$ on the other hand.
This relation will be crucial in our analysis.
We note that \eqref{eqn:Master Final} matches the effective potential derived from the ten-dimensional Einstein equations in \cite{Giddings:2005ff}, see e.g.~equation (5.30) of \cite{Giddings:2005ff}.

An equivalent route to deriving \eqref{eqn:Master Final} is to first follow the steps leading to the Einstein-minus-Bianchi equation (2.30) of \cite{Giddings:2001yu}, which in our conventions reads
\begin{equation}\label{ourgkp}
\nabla^2\Phi_-=e^{-4A}\diff_a \Phi_-\diff^a\Phi_-+\frac{1}{2}\kappa_{10}^2e^{2A+2u}\bigl(\hat{T}_{ab}G^{ab}-\hat{T}_{\mu\nu}G^{\mu\nu}\bigr)-2\kappa_{10}^2\mu_3 e^{8A-4u}\rho_{D3}+e^{8u}\mathcal{R}_4^{\mathrm{Ref.[12]}}\,.
\end{equation}
Because we have made explicit the breathing mode $u$, which was instead implicit in the metric ansatz of \cite{Giddings:2001yu}, the scalar curvatures there and here are related by
\begin{equation}\label{rshift}
\mathcal{R}_4^{\mathrm{Ref.[12]}}=\mathcal{R}_4[g]+12\,\square u -24\diff_\mu u\diff^\mu u\,.
\end{equation}
Substituting \eqref{rshift} in \eqref{ourgkp} and using the Einstein equations and Bianchi identity, one arrives at \eqref{eqn:Master Final}.
The point we would like to stress is that equation (2.30) of \cite{Giddings:2001yu} --- which has been the basis of a number of constraints on compact solutions --- and the master equation \eqref{eqn:Master Final} contain equivalent information, \emph{provided} that one correctly accounts for the breathing mode as in \eqref{rshift}.

\section{Stress-energy of Gaugino Condensate} \label{sec:SUSY}

Our goal is to examine the de Sitter scenario of \cite{KKLT} using the ten-dimensional equations of motion.
In the four-dimensional effective theory, the scalar potential has two components: an F-term potential for the moduli of an $\mathcal{N}=1$ supersymmetric compactification, and a supersymmetry-breaking contribution from one or more anti-D3-branes.  We will examine these in turn: in this section we consider the ten-dimensional configuration without anti-D3-branes, and then in \S\ref{sec:dS} we incorporate the effects of anti-D3-branes.

The relevant moduli at low energies are the K\"ahler moduli of the Calabi-Yau orientifold $M$, because the complex structure moduli and axiodilaton acquire mass from $G_3$ flux at a higher scale.\footnote{If D3-branes are present, their position moduli have masses parametrically comparable to those of the K\"ahler moduli, and the corresponding potential can be computed in ten dimensions \cite{Baumann:2010sx}: see Appendix \ref{app:d3}.}  For simplicity of presentation we will consider a single K\"ahler modulus, which we denote by $T$,
but our method applies more generally.

The four-dimensional analysis of \cite{KKLT} established that in the presence of a suitably small\footnote{The statistical approach of Denef and Douglas \cite{douglas:distributions} gives strong evidence that (in the spirit of \cite{polchinski:anthropics}) one can fine-tune the classical flux superpotential $W_0 = \langle W_{\mathrm{flux}}\rangle$ to be small.
This conclusion is supported by \cite{sk:awful}, which explicitly demonstrates that values of $W_0$ small enough for control of the instanton expansion are achievable even with few complex structure moduli.}
classical flux superpotential, combined with a nonperturbative superpotential from Euclidean D3-branes or from gaugino condensation on D7-branes, the K\"ahler modulus $T$ is stabilized in an $\mathcal{N}=1$ supersymmetric $AdS_4$ vacuum.
To recover this result from ten dimensions, we need to understand how these two superpotential terms  correspond to ten-dimensional field configurations.

First of all, the Gukov-Vafa-Witten flux superpotential \cite{Gukov:1999ya}
\begin{equation}
W_{\mathrm{flux}}=\pi\int G\wedge \Omega \label{eq:GVW}
\end{equation}
encodes in the four-dimensional effective theory the interaction corresponding to the term
\begin{equation}\label{sflux}
S_{\mathrm{flux}}=-\frac{1}{2\kappa_{10}^2}\int \mathrm{d}^{10} X \sqrt{-G}\,\frac{G_3\cdot\overbar{G}_3}{2\,\mathrm{Im}\,\tau}
\end{equation}
in the ten-dimensional action \eqref{sfullis}.
In particular, the ten-dimensional stress-energy associated to $W_{\mathrm{flux}}$ is that computed from \eqref{sflux}.

In the remainder of this section, we will describe
the gaugino condensate superpotential in similarly ten-dimensional terms, and compute the contribution $T_{\mu\nu}^{\langle\lambda\lambda\rangle}$  of gaugino condensation on D7-branes to the ten-dimensional stress-energy tensor. We will see that the stress energy $T_{\mu\nu}^{\langle\lambda\lambda\rangle}$  arises from gaugino-flux couplings generalizing those derived by C\'amara, Ib\'a\~{n}ez, and Uranga in \cite{Camara:2004jj}, and also from associated nonsingular four-gaugino terms.
We will then show that this stress-energy\footnote{In Appendix \ref{app:d3} we account for the terms other than $T_{\mu\nu}^{D7}$ in \eqref{eqn:Master Final}, and demonstrate that our conclusions remain unchanged.} leads to a potential for the K\"ahler modulus that exactly matches the F-term potential of \cite{KKLT}.

Because the gaugino condensate relies on the dynamics of the D7-brane gauge theory below the Kaluza-Klein scale, it is not entirely obvious that a ten-dimensional description of gaugino condensation should exist at all.
However, as explained in \cite{Baumann:2010sx}, one can consider D7-branes wrapping a divisor that is very small compared to the entire compact space.
A localized `observer' far from the D7-branes, such as a distant D3-brane, should then be able to treat them as a fuzzy source.
This approach turns out to be fruitful: we will exhibit below a precise correspondence between the ten-dimensional and four-dimensional computations of the potential for the K\"ahler modulus, just as the four-dimensional result for the potential of a D3-brane probe was obtained from ten dimensions in \cite{Baumann:2010sx}.\footnote{See Appendix \ref{app:d3} for a computation of the D3-brane potential that extends the result of \cite{Baumann:2010sx}.}

\subsection{Four-dimensional effective theory}

We begin by recalling results from the four-dimensional effective theory that we aim to recover from ten dimensions.
Dimensional reduction of the theory on a stack of D7-branes wrapping a divisor $D$ leads at low energies, and in the limit that gravity decouples, to the $\mathcal{N}=1$ supersymmetric Yang-Mills Lagrangian density
\begin{equation}\label{ymaction}
\frac{1}{16\pi i}\int d^2\theta f(T)\, W_\alpha W^\alpha +c.c.\,,
\end{equation}
where we have adopted the conventions of \cite{Terning:2006bq}, but suppress
Lie algebra indices.  We will denote the dual Coxeter number of the gauge group by $N_c$.

Classically, the $\mathcal{N}=1$ supergravity theory associated to \eqref{ymaction}, for D7-branes in a background whose moduli potential is described by a classical flux superpotential $W_{\mathrm{flux}}$, has the Lagrangian density (see e.g.~\cite{Kallosh:2019oxv})
\begin{align}
\mathcal{L}=&-\frac{1}{4}\text{Re}f(T)F_{\mu\nu}F^{\mu\nu}-i\bar{\lambda}\bar{\sigma}^\mu\diff_\mu\lambda\,\text{Re}f(T)-\frac{1}{4}\lambda\lambda\, e^{\kappa_4^2 K(T,\overbar{T})/2} K^{T\overbar{T}}\partial_T f(T) K_{\overbar{T}}\overline{W}_{\mathrm{flux}}+c.c.\nonumber\\
&+\frac{3\kappa_4^2}{64}\Bigl(\bar{\lambda}\bar{\sigma}^\mu\lambda\,\text{Re}f(T)\Bigr)^2-\frac{1}{16}\lambda\lambda\bar{\lambda}\bar{\lambda} K^{T\overbar{T}}\partial_Tf(T) \partial_{\overbar{T}}\bar{f}(\overbar{T})\,,\label{eqn:gauge action}
\end{align}
which reduces to \eqref{ymaction} in the limit $\kappa_4 \to 0$. We take the divisor $D$ to be rigid, so that the Yang-Mills theory has no charged matter.
Here, the D7-brane gauge coupling is given by the holomorphic expression\footnote{The normalization $f(T)=T/(2\pi)$ was used in the study of gaugino-flux couplings in \cite{Dymarsky:2010mf,Moritz:2017xto}, but we
take instead $f(T)=T/(4\pi)$ for ease of comparison to the supergravity literature.}
\begin{equation}\label{deftClassical}
f(T)=\frac{T}{4\pi} \qquad \mathrm{with}~T:=\int_D \sqrt{g_6} e^{-4A+4u} +i\int_D C_4\,.
\end{equation}
However, as explained by \cite{Kaplunovsky:1994fg}, in a quantum mechanical effective field theory treatment of supergravity (as opposed to classical supergravity) the gauge coupling function receives a non-holomorphic contribution from the K\"ahler potential:
\begin{equation}\label{deft}
f(T,\bar T)=\frac{T}{4\pi} - \frac{N_c}{16\pi^2} \kappa_4^2 K(T,\bar T) \,.
\end{equation}
This term is present thanks to an anomaly in the Weyl rescaling that transforms fields from the normalization which has linearly realized supersymmetry (when one restores the auxiliary fields) and a holomorphic $f(T)$ to the physical normalization employed in (3.4). Because of this, the usual expression for the gaugino bilinear expectation value in terms of the gauge coupling function depends non-holomorphically on $T$ \cite{Kaplunovsky:1994fg}:
\begin{equation}\label{eqn:gaugino vev}
\langle \lambda\lambda \rangle = - \frac{32\pi^2}{N_c} \mathcal{A}\, e^{-\frac{8\pi^2}{N_c} f(T,\bar T)} \,.
\end{equation}
Similarly, the classical Lagrangian \eqref{eqn:gauge action} requires a number of modifications to account for the fact that $f$ is not holomorphic.

After integrating out the vector multiplet, one obtains an effective field theory valid below the confinement scale that involves only the chiral superfield containing the K\"ahler modulus and the supergravity multiplet. This has the superpotential
\begin{equation}\label{wis}
W = W_{\mathrm{flux}} + W_{\mathrm{np}}\,,
\end{equation}
where \cite{Kaplunovsky:1994fg}
\begin{equation}\label{wnpis}
W_{\mathrm{np}}=\mathcal{A}\,e^{-\frac{2\pi T}{N_c}}\,.
\end{equation}
This leads to the relation
\begin{equation}\label{wnpll}
\langle \lambda\lambda \rangle = - \frac{32\pi^2}{N_c} e^{\kappa_4^2 K/2} W_{\mathrm{np}} \,.
\end{equation}
The Pfaffian prefactor $\mathcal{A}$ depends on the complex structure moduli and the positions of any D3-branes: see \cite{Berg:2004ek,Baumann:2006th}. Finally, the K\"ahler potential is\footnote{Although the complex structure moduli and dilaton receive supersymmetric masses from the flux background, we retain the associated terms in \eqref{kpotis} because their expectation values matter for the overall normalization.
The K\"ahler potential \eqref{kpotis} is consistent with that of \cite{DeWolfe:2002nn,Koerber:2007xk,Kallosh:2000ve} --- see Appendix \ref{app:conv} for details of our conventions.}
\begin{equation}\label{kpotis}
K=-3\log\bigl(T+\overbar{T}\bigr)-\log\bigl(-i(\tau-\overbar{\tau})\bigr)-\log\left(i\int_M e^{-4A}\Omega\wedge\overline{\Omega} \right)+\log\Bigl(2^7 \mathcal{V}^3\Bigr)\,.
\end{equation}
In terms of these functions, the F-term potential in this effective field theory is
\begin{equation}\label{4dpot}
V = e^{\kappa_4^2 K}\left(K^{T\overbar{T}} D_T W D_{\overbar{T}}\overline{W}-3\kappa_4^2 W\overline{W} \right)\,.
\end{equation}

Our goal is now to show that the F-term potential \eqref{4dpot}, which we have just recalled as a result in four-dimensional supergravity,
can also be derived from the \emph{ten-dimensional} equations of motion, upon assigning the vev \eqref{eqn:gaugino vev} and examining the ten-dimensional stress-energy.

\subsection{D7-brane gaugino couplings}

Now we turn to ten dimensions. To describe the backreaction of the gaugino condensate on the bulk fields, we must relax the Calabi-Yau condition and employ generalized complex geometry, as in \cite{Grana:2005sn,Benmachiche:2006df,Koerber:2007xk,Koerber:2010bx}.
In particular, as reviewed in Appendix \ref{app:conv}, the single covariantly constant spinor is replaced by two internal Killing spinors $\eta_1$ and $\eta_2$.
We can combine these to form a bispinor $\Phi_1$, defined as
\begin{equation}
\Phi_1:=-\frac{8i}{|\eta|^2}\eta_1\otimes \eta_2^\dagger\,,
\end{equation}
and we also define
\begin{equation}
\mathfrak{t}:=\mathrm{Re}\Bigl(e^{-\phi+(\phi/4-A)\hat{p}}\Phi_1\Bigr)\,,
\end{equation}
where the operator $\hat{p}$ is defined by
\begin{equation}
\hat{p}\,C_p:= p\,C_p
\end{equation}
for a $p$-form $C_p$ \cite{Heidenreich:2010ad}.
In type IIB string theory compactified on an orientifold of a Calabi-Yau threefold, and in the absence of nonperturbative effects, one has $\mathfrak{t}=0$.
However, upon including the effects of gaugino condensation, $\mathfrak{t}$ develops a nonvanishing two-form component \cite{Dymarsky:2010mf}, cf.~\eqref{eqn:mathfrakt}, that will be important for our analysis.

We now study the action of D7-branes on such a generalized complex geometry. The eight-dimensional action describing a stack of D7-branes is derived in Appendix \ref{app:conv} via dimensional reduction and T-dualization of the type I action. We will highlight the important changes that occur when, instead of dimensionally reducing these D7-branes on a divisor in a Calabi-Yau orientifold, one wraps a divisor in a generalized complex geometry. Our findings reproduce results of \cite{Koerber:2007xk}.

\subsubsection{Gaugino-flux couplings}

To write the flux superpotential and the gaugino-flux couplings on D7-branes in a generalized complex geometry, we first define
\begin{equation}\label{frakGdef}
\mathfrak{G}:= G_3+i\mathrm{d}\mathfrak{t}\,,
\end{equation} and
\begin{equation}\label{frakG2def}
\mathfrak{G}^{[2]}:= G_3+i\mathrm{d}_2\mathfrak{t}\,,
\end{equation} where $\mathrm{d}_2$ is a differential operator defined in terms of coordinates along the D7-brane, and is given in Appendix \ref{decompapp}.

The gaugino-flux couplings on D7-branes are determined by the supersymmetric Born-Infeld action.
In the conventions of \cite{Lust:2008zd,Dymarsky:2010mf}, with the
metric ansatz \eqref{eqn:metric ansatz}, and recalling that we have set $(2\pi)^2\alpha'=1$, these couplings -- on a divisor \emph{in a Calabi-Yau orientifold}, not a generalized complex geometry -- are
\begin{equation}\label{ciunogcg}
S_{G\lambda\lambda}=\frac{i}{32\pi}\int \sqrt{-g_4\,g_6}e^{\phi/2}e^{-2u}\,G_3\cdot \Omega\,\bar{\lambda}\bar{\lambda}\,\delta^{(0)}+c.c.
\end{equation}
We re-derive this interaction via dimensional reduction of the eight-dimensional D7-brane action in Appendix \ref{app:conv}.

In similar fashion, we find the action that one obtains from wrapping a divisor in a generalized complex geometry. The details are relegated to Appendix \ref{app:conv}; the result, in agreement with \cite{Benmachiche:2006df,Koerber:2007xk}, is that one should promote\footnote{Discussions of \eqref{thepromotion} in this context include \cite{LMtalk} and the recent work \cite{Bena:2019mte}.}
\begin{equation}\label{thepromotion}
G_3 \to G_3+i\mathrm{d}_2\mathfrak{t} \equiv \mathfrak{G}^{[2]}\,.
\end{equation}
Thus, \eqref{ciunogcg} becomes (cf.~\cite{Dymarsky:2010mf})
\begin{equation}\label{eqn:ciugcg}
S_{\mathfrak{G}\lambda\lambda}=\frac{i}{32\pi}\int \sqrt{-g_4\,g_6}e^{\phi/2}e^{-2u}\, \mathfrak{G}^{[2]} \cdot \Omega\, \bar{\lambda}\bar{\lambda}\, \delta^{(0)}+c.c.
\end{equation}

One can likewise generalize the familiar flux superpotential \eqref{eq:GVW}.
The superpotential in a generalized complex geometry has been studied in e.g.~\cite{Grana:2005ny,Benmachiche:2006df,Micu:2007rd,Koerber:2007xk,Lust:2008zd} from several angles, for example by computing the mass and supersymmetry transformation of the gravitino.
In our class of solutions this superpotential takes the form
\begin{equation} \label{eqn:flux susy}
W_{\mathrm{GCG}} = \pi \int_M \mathfrak{G}\wedge\Omega \,,
\end{equation} as we show in Appendix \ref{wgravapp}.

To relate $W_{\mathrm{GCG}}$ to the superpotential $W = W_{\mathrm{flux}} + W_{\mathrm{np}}$ given in \eqref{wis}, we impose the ten-dimensional Killing spinor equations that govern supersymmetric solutions.
In Appendix \ref{wgravapp} we show that
\begin{equation} \label{eqn:wgcgvev}
\langle W_{\mathrm{GCG}} \rangle = W_{\mathrm{flux}} + W_{\mathrm{np}}\,,
\end{equation} where the brackets indicate evaluation in the supersymmetric configuration.
The generalized complex geometry thus elegantly encodes the effects of the nonperturbative superpotential.

Evaluating the gaugino-flux coupling \eqref{eqn:ciugcg}, one finds (see Appendix \ref{appto4d} for details of the computation)
\begin{align}
S_{\mathfrak{G}\lambda\lambda}=-\kappa_4^2 \int_X \sqrt{-g_4}e^{\kappa_4^2 K} K^{T\overbar{T}} \partial_{\overbar{T}} \overline{W} K_T W +c.c.+S_{\lambda\lambda}^{\mathrm{sing}}\,,\label{eqn:ciugcg dim reduction}
\end{align}
where $S_{\lambda\lambda}^{\mathrm{sing}}$ is a singular contribution, given in \eqref{ssingis}, that is treated in Appendix \ref{app:d3}.

\subsubsection{Four-gaugino coupling}

We similarly demonstrate in Appendix \ref{app:conv}, by dimensional reduction and T-dualization of the ten-dimensional type I action, that there is a four-gaugino coupling\footnote{The importance of four-gaugino couplings in this context was stressed in \cite{Hamada:2018qef}.} on D7-branes given by
\begin{equation}\label{eqn:fourgaugino}
S_{\lambda\lambda\lambda\lambda}=-\frac{1}{6144\pi^3}\int \sqrt{-g_4\,g_6}e^{-4A+8u} \nu\, \Omega\cdot\overline{\Omega}\, |\lambda\lambda|^2\,\delta^{(0)},
\end{equation}
where $\nu \equiv \mathcal{V}_\perp^{-1} =\mathcal{V}^{-1}\int_D\sqrt{g_6} e^{-4A}$ is the inverse of the volume $\mathcal{V}_\perp$ transverse to the D7-branes. Upon assigning the gaugino bilinear vev \eqref{eqn:gaugino vev}, the four-gaugino term \eqref{eqn:fourgaugino} dimensionally reduces to
\begin{align}\label{eqn:fourgauginofinal}
S_{\lambda\lambda\lambda\lambda}=-\int_X\sqrt{-g_4} e^{\kappa_4^2K}K^{T\overbar{T}}\diff_T W\diff_{\overbar{T}}\overline{W}.
\end{align}
See Appendix \ref{app:conv} for details of the computation.

\subsection{Ten-dimensional stress-energy}

We can now obtain the F-term potential for the K\"ahler modulus $T$ from the ten-dimensional field configuration.
Upon assigning the gaugino bilinear vev \eqref{eqn:gaugino vev} and using \eqref{eqn:ciugcg dim reduction}, the properly-holomorphic  gaugino-flux coupling \eqref{eqn:ciugcg} evaluates to
 \begin{equation}\label{eqn:gcgciuvev}
\mathcal{L}_{\mathfrak{G}\lambda\lambda}= -\kappa_4^2 e^{\kappa_4^2K}K^{T\overbar{T}}\Bigl(\diff_T W K_{\overbar{T}}\overline{W}+c.c.\Bigr)+\mathcal{L}_{\lambda\lambda}^{\mathrm{sing}}\,.
\end{equation}
The associated ten-dimensional stress-energy is
\begin{equation}\label{tmunu2lambda}
T_{\mu\nu}^{\lambda\lambda}:= -\frac{2}{\sqrt{G}}\frac{\delta \mathcal{L}_{\mathfrak{G}\lambda\lambda}}{\delta G^{\mu\nu}}= \frac{i}{32\pi}e^{4A+\phi/2-2u}\,\mathfrak{G}^{[2]}\cdot \Omega\,\bar{\lambda}\bar{\lambda}\,\delta^{(0)}g_{\mu\nu}+c.c.\,,
\end{equation}
which integrates to
\begin{equation}\label{eqn:aterm}
-\int_M \sqrt{g_6}e^{-4A}T_{\mu\nu}^{\lambda\lambda}g^{\mu\nu}=4\kappa_4^2 e^{\kappa_4^2K} K^{T\overbar{T}}\partial_T W K_{\overbar{T}}\overline{W}+c.c.-4S_{\lambda\lambda}^{\mathrm{sing}}.
\end{equation}
Setting aside $S_{\lambda\lambda}^{\mathrm{sing}}$ for the moment, we see from \eqref{eqn:aterm} that the gaugino-flux coupling contributes a finite term in the F-term potential for the K\"ahler modulus $T,$
\begin{equation}\label{twogauginoV}
V_{\lambda\lambda}= \kappa_4^2 e^{\kappa_4^2K} K^{T\overbar{T}}\partial_T W K_{\overbar{T}}\overline{W}+c.c.
\end{equation}
We now follow the same steps for the four-gaugino coupling.
From \eqref{eqn:fourgaugino}, $T_{\mu\nu}^{\lambda\lambda\lambda\lambda}$ is
\begin{equation}\label{tmunu4lambda}
T_{\mu\nu}^{\lambda\lambda\lambda\lambda}:= -\frac{2}{\sqrt{G}}\frac{\delta \mathcal{L}_{\lambda\lambda\lambda\lambda}}{\delta G^{\mu\nu}}=-e^{8u}
\frac{\nu\,\Omega\cdot\overline{\Omega}}{6144\pi^3}\,|\lambda\lambda|^2\, \delta^{(0)} g_{\mu\nu}\,,
\end{equation}
which integrates to
\begin{equation}\label{eqn:fourfermiVterm}
-\int_M \sqrt{g_6}e^{-4A}T_{\mu\nu}^{\lambda\lambda\lambda\lambda} g^{\mu\nu}=4e^{\kappa_4^2 K}K^{T\overbar{T}}\diff_T W\diff_{\overbar{T}}\overline{W}.
\end{equation}
The four-gaugino coupling \eqref{eqn:fourgaugino} therefore contributes the term
\begin{equation}\label{fourgauginoV}
V_{\lambda\lambda\lambda\lambda} = e^{\kappa_4^2 K}K^{T\overbar{T}} \diff_T W\diff_{\overbar{T}} \overline{W}\,.
\end{equation}
The total ten-dimensional stress-energy is then
\begin{equation}\label{totalTmunu}
T_{\mu\nu}^{\langle\lambda\lambda\rangle}:=T_{\mu\nu}^{\lambda\lambda}+T_{\mu\nu}^{\lambda\lambda\lambda\lambda}\,,
\end{equation} with $T_{\mu\nu}^{\lambda\lambda}$ given by \eqref{tmunu2lambda} and with $T_{\mu\nu}^{\lambda\lambda\lambda\lambda}$ given by \eqref{tmunu4lambda}.
Combining \eqref{twogauginoV} and \eqref{fourgauginoV} to evaluate
the integral of $T_{\mu\nu}^{\langle\lambda\lambda\rangle}$ over the internal space, and continuing to set aside the singular term $S_{\lambda\lambda}^{\mathrm{sing}}$, we conclude that the ten-dimensional field configuration sourced by gaugino condensation on D7-branes gives rise to the four-dimensional scalar potential,
\begin{equation}\label{10dans}
\boxed{\vphantom{\Biggl(\Biggr)} V = e^{\kappa_4^2 K}\Bigl(K^{T\overbar{T}} D_T W D_{\overbar{T}}\overline{W}-3\kappa_4^2 W\overline{W} \Bigr)}\,,
\end{equation}
and so precisely recovers the potential \eqref{4dpot} computed in four-dimensional supergravity.
In summary, we have shown that the \emph{ten-dimensional} equation of motion \eqref{eqn:Master Final}, incorporating the stress-energy
$T_{\mu\nu}^{\langle\lambda\lambda\rangle}$ in \eqref{totalTmunu},
requires that the Einstein-frame scalar curvature $\mathcal{R}_4[g]$ takes exactly the value demanded by the \emph{four-dimensional} Einstein equation \eqref{ee4} with the scalar potential \eqref{4dpot}, i.e.~the value computed in the four-dimensional effective theory in \cite{KKLT}.  This is one of our main results.

Before proceeding, we will comment briefly on the singularities in our solution, including $S_{\lambda\lambda}^{\mathrm{sing}},$ deferring a complete treatment to Appendix \ref{app:d3}.
The ten-dimensional configuration corresponding to gaugino condensation on D7-branes contains specific singular field profiles, because the D7-branes are localized to a complex hypersurface in the internal space.
In particular, as shown in \cite{Baumann:2010sx}, the $G_-$ flux sourced by gaugino condensation is
\begin{equation}
(G_-)_{a\bar{c}\bar{d}}=-ie^{-4A-\phi/2+8u}\frac{\lambda\lambda}{32\pi^2}\diff_a\diff_b G_{(2)}(z;z_{D7})g^{b\bar{b}}\overline{\Omega}_{\bar{b}\bar{c}\bar{d}}\,,\label{eqn:flux from gauginofirst}
\end{equation} where $G_{(2)}$ is the Green's function on the internal space transverse to the D7-branes, with complex coordinate $z$.
Similarly, it was shown in \cite{Dymarsky:2010mf} that gaugino condensation sources $G_+$ flux that is localized on the D7-branes:
\begin{equation}
G_+=- i\frac{e^{-4A}}{64\pi^2}e^{-\phi/2}\,\lambda\lambda\,\overline{\Omega}\,\delta^{(0)}.\label{eqn:localized flux}
\end{equation}
Upon evaluating the D7-brane action and the bulk supergravity action in the presence of these flux configurations, one finds divergent contributions to the stress-energy, and in turn to the six-dimensional curvature $\mathcal{R}_6$ in \eqref{eqn:Master Final}.
However, we show explicitly in Appendix \ref{app:d3} that a highly nontrivial cancellation occurs:
\emph{all the divergences appearing in \eqref{eqn:Master Final} cancel}, and the finite piece that remains gives exactly \eqref{10dans}.

\section{Anti-D3-branes and Gaugino Condensation} \label{sec:dS}

Thus far we have shown that the F-term potential in and around the $\mathcal{N}=1$ supersymmetric $AdS_4$ vacuum of \cite{KKLT} can be obtained in two ways.  The first is four-dimensional supergravity, as originally argued in \cite{KKLT}.  The second derivation, as shown above, is from ten-dimensional supergravity, supplemented with the gaugino bilinear vev \eqref{eqn:gaugino vev} substituted into the two-gaugino and four-gaugino terms in the D7-brane action.

We now turn to the effects of anti-D3-branes, and to the study of four-dimensional de Sitter vacua from ten dimensions.

\subsection{Decompactification from anti-D3-branes}\label{d3sub}

We first consider the effects of an anti-D3-brane in a no-scale flux compactification, \emph{without} a nonperturbative superpotential for the K\"ahler moduli.

The Dirac-Born-Infeld action of a spacetime-filling anti-D3-brane at position $y_{\overbar{D3}}$ in the internal space leads to
the stress-energy tensor
\begin{equation}\label{eqn:stress energy D3}
T_{\mu\nu}^{\overbar{D3}}=-\mu_3e^{8A-12u}g_{\mu\nu}\delta(y-y_{\overbar{D3}})\,.
\end{equation}
Inserting \eqref{eqn:stress energy D3} in \eqref{eqn:Master 2}, we learn that including a single anti-D3-brane in a no-scale background leads to a shift in the effective potential,\footnote{As explained in \cite{Kachru:2003sx}, if the anti-D3-brane is in a strongly warped region, the dependence on the breathing mode becomes $e^{-8u}$ rather than $e^{-12u}$.}
\begin{align}
\frac{1}{4}M_{\mathrm{pl}}^2 \delta \mathcal{R}_4[g]= 2\mu_3 e^{-12u}e^{4A}(y_{\overbar{D3}})\,.\label{eqn:anti brane energy}
\end{align}
The potential energy captured by \eqref{eqn:anti brane energy} is minimized in the infinite volume limit $u\rightarrow \infty,$ so in the absence of any other effects an anti-D3-brane will cause runaway decompactification.  The expression \eqref{eqn:anti brane energy} agrees with the four-dimensional analysis of \cite{KKLT}.

\subsection{Interactions of anti-D3-branes and gaugino condensation}

To examine the ten-dimensional stress-energy, we write the ten-dimensional field configuration in the schematic form
\begin{equation}
\phi = \phi_{\mathrm{bg}} + \delta\phi\,,
\end{equation}
with
\begin{equation}
\delta\phi =  \delta\phi\vert_{\langle\lambda\lambda\rangle} + \delta\phi\vert_{\overbar{D3}}\,.
\end{equation}
Here $\phi$ is any of the ten-dimensional fields, $\phi_{\mathrm{bg}}$ is the field configuration when neither gaugino condensation nor anti-D3-branes are included as sources, $\delta\phi\vert_{\langle\lambda\lambda\rangle}$ is the change in the field configuration when gaugino condensation is included as a source, and $\delta\phi\vert_{\overbar{D3}}$ is the change in the field configuration when $p$ anti-D3-branes are included as a source.

The changes $\delta\phi\vert_{\langle\lambda\lambda\rangle}$ and $\delta\phi\vert_{\overbar{D3}}$ are each parametrically small away from their corresponding sources: $\langle\lambda\lambda\rangle$ is exponentially small by dimensional transmutation,
and the anti-D3-brane is in a warped region.  Because the anti-D3-branes and the D7-brane stack are widely-separated, we can safely neglect the nonlinear corrections to the field configuration resulting from simultaneously including both gaugino condensation and anti-D3-branes as sources.\footnote{See \cite{Gandhi:2011id} and Appendix \ref{app:spec} for further details and references on nonlinear interactions.}

Separating the ten-dimensional Lagrange density as
\begin{equation}\label{Lsplit}
\mathcal{L} = \mathcal{L}_{\mathrm{SUSY}} + p\,\mathcal{L}_{\mathrm{loc}}^{\overbar{D3}}\,,
\end{equation} with $\mathcal{L}_{\mathrm{SUSY}}=\mathcal{L}_{\mathrm{bulk}} + \mathcal{L}_{\mathrm{loc}}^{D7}$,
the total ten-dimensional stress-energy can be written
\begin{equation}\label{thetotT}
T_{\mu\nu} =-\frac{2}{\sqrt{-G}} \frac{\delta\mathcal{L}_{\mathrm{SUSY}}}{\delta G^{\mu\nu}}-\frac{2}{\sqrt{-G}}\frac{\delta\mathcal{L}_{\mathrm{loc}}^{\overbar{D3}}}{\delta G^{\mu\nu}} \equiv T_{\mu\nu}^{\langle\lambda\lambda\rangle}\Bigr\vert_{\phi}+
T_{\mu\nu}^{\overbar{D3}}\Bigr\vert_{\phi}\,,
\end{equation}
which we write as
\begin{equation}\label{deftint}
T_{\mu\nu} = T_{\mu\nu}^{\langle\lambda\lambda\rangle}\Bigr\vert_{\phi_{\mathrm{bg}}+\delta\phi\vert_{\langle\lambda\lambda\rangle} }  + p\,T_{\mu\nu}^{\overbar{D3}}\Bigr\vert_{\phi_{\mathrm{bg}}}  + T_{\mu\nu}^{\mathrm{int}}\,.
\end{equation}
The first  term on the right in \eqref{deftint} is the stress-energy \eqref{totalTmunu} of gaugino condensation on D7-branes,
computed in the field configuration $\phi = \phi_{\mathrm{bg}} + \delta\phi\vert_{\langle\lambda\lambda\rangle}$, i.e.~without including the backreaction of any anti-D3-branes, as in \S\ref{sec:SUSY}.
The second term is the stress-energy \eqref{eqn:stress energy D3} due to the Dirac-Born-Infeld action of $p$ anti-D3-branes, computed as probes of the background $\phi = \phi_{\mathrm{bg}}$, as in \S\ref{d3sub}.

The interaction term $T_{\mu\nu}^{\mathrm{int}}$ is \emph{defined} by \eqref{deftint}, and
captures the stress-energy due to the interactions of the anti-D3-branes and the condensate: specifically, the correction to $T_{\mu\nu}^{\langle\lambda\lambda\rangle}$ from the shift $\delta\phi\vert_{\overbar{D3}}$, and the correction to $T_{\mu\nu}^{\overbar{D3}}$ from the shift $\delta\phi\vert_{\langle\lambda\lambda\rangle}$.\footnote{Corrections to $T_{\mu\nu}^{\overbar{D3}}$ from the shift $\delta\phi\vert_{\overbar{D3}}$ are subleading.}
We will now explain why $T_{\mu\nu}^{\mathrm{int}}$ can be neglected, so that $T_{\mu\nu}$ is well-approximated by the first two terms on the right in \eqref{deftint}.
Since we have already shown in \S\ref{sec:SUSY} and \S\ref{d3sub} that these two terms together precisely reproduce the four-dimensional effective potential of \cite{KKLT}, establishing that $T_{\mu\nu}^{\mathrm{int}}$ is negligible
will complete our demonstration that the ten-dimensional equations of motion recover the result of \cite{KKLT}.

To show that the interaction $T_{\mu\nu}^{\mathrm{int}}$ is negligible,
one can consider the leading effects of $p$ anti-D3-branes on the ten-dimensional fields at the location of the the D7-branes, and evaluate the resulting correction to the ten-dimensional stress-energy $T_{\mu\nu}^{\langle\lambda\lambda\rangle}$.

As a cross-check, one can reverse the roles of source and probe, estimate the leading effects of the D7-brane gaugino condensate on the ten-dimensional fields at the location of the anti-D3-branes, and evaluate the resulting correction to the stress-energy $p\,T_{\mu\nu}^{\overbar{D3}}$ computed from the probe action of $p$ anti-D3-branes.

The methodology for the computation is parallel in the two cases, and builds on investigations of supergravity solutions sourced by antibranes \cite{Kachru:2002gs,DeWolfe:2008zy,
McGuirk:2009xx,Bena:2009xk,Bena:2010ze,Bena:2011hz,Dymarsky:2011pm,Bena:2011wh,Cohen-Maldonado:2015ssa,Aalsma:2018pll,Armas:2018rsy}, and of D3-brane potentials in warped throats \cite{Kachru:2003sx,Baumann:2006th,DeWolfe:2007hd,Baumann:2008kq,Baumann:2010sx,Gandhi:2011id}.
One can approximate the Klebanov-Strassler throat as a region in $AdS_5 \times T^{1,1}$, and use the Green's functions for the conifold (see e.g.~\cite{Klebanov:2007us}) to compute the influence of a localized source --- i.e.,~the anti-D3-branes or the D7-brane gaugino condensate --- on distant fields.
Far away from the source, the dominant effects appear as certain leading multipoles, corresponding to the lowest-dimension operators to which the source couples.  Schematically (see Appendix \ref{app:spec} for details),
\begin{equation}\label{genpertdelta}
\delta\phi = \sum_{\Delta} \alpha_{\Delta} \Bigl(\frac{r}{r_{\mathrm{UV}}}\Bigr)^{-\Delta} + \beta_{\Delta} \Bigl(\frac{r}{r_{\mathrm{UV}}}\Bigr)^{\Delta-4}\,,
\end{equation} where $\Delta$ is the dimension of an operator $\mathcal{O}_{\Delta}$ in the dual field theory, $r$ is the radial coordinate of the throat, and $r_{\mathrm{UV}}$ is the location of the ultraviolet end of the throat.  The coefficients $\alpha_{\Delta}$ and $\beta_{\Delta}$ correspond to expectation values and sources, respectively, for the dual operator.

The spectrum of operators of the Klebanov-Witten theory \cite{Klebanov:1998hh} dual to $AdS_5 \times T^{1,1}$ is well-understood, due to the pioneering work of Gubser \cite{Gubser:1998vd} and of Ceresole et al.~\cite{Ceresole:1999ht,Ceresole:1999zs} (see also
\cite{Aharony:2005ez,Baumann:2010sx,Gandhi:2011id,gandhisjors}), and moreover there are many quantitative cross-checks of the long-distance solutions created by anti-D3-branes \cite{DeWolfe:2008zy,Kachru:2007xp,Baumann:2008kq,Bena:2009xk,Bena:2010ze,Dymarsky:2011pm,Bena:2011wh,Baumann:2010sx,Berg:2010ha,Dymarsky:2013tna}
and by gaugino condensates \cite{Baumann:2008kq,Baumann:2010sx,Heidenreich:2010ad,Dymarsky:2010mf,Moritz:2017xto}.
In Appendix \ref{app:spec} we assemble key results from this literature, and then apply them to compute the leading interactions of anti-D3-branes with a gaugino condensate.
A brief summary is as follows.

In the linearized supergravity solution sourced by anti-D3-brane backreaction, as in \cite{Bena:2009xk,Bena:2010ze,Bena:2011hz,Dymarsky:2011pm}, the leading effects of anti-D3-branes in the infrared on the D7-brane gaugino condensate are mediated by expectation values for operators of dimension $\Delta \ge 8$, cf.~\eqref{d3bargives1},\eqref{o8is}, and so can be neglected when the hierarchy of scales in the throat is large.  Nonlinear effects are likewise negligible \cite{Gandhi:2011id,McAllister:2016vzi}.

Similarly, in the supergravity solution sourced by gaugino condensate backreaction, the leading effects of the D7-brane gaugino condensate on the anti-D3-branes are negligible compared to the probe anti-D3-brane action in the Klebanov-Strassler background, cf.~\eqref{delvfromall},\eqref{delvfromallnonlin} \cite{Gandhi:2011id,gandhisjors}, both at the linear and the nonlinear level.

In sum,
the dominant influence of the anti-D3-branes on the gaugino condensate is via the breathing mode $e^u$.
All other interactions are suppressed by further powers of the warp factor.
We have therefore established that
\begin{equation}\label{thetotTapproxyes}
T_{\mu\nu} \approx T_{\mu\nu}^{\langle\lambda\lambda\rangle} + p\,T_{\mu\nu}^{\overbar{D3}}+\ldots\,,
\end{equation} where $T_{\mu\nu}^{\langle\lambda\lambda\rangle}$ is given by \eqref{totalTmunu}, $T_{\mu\nu}^{\overbar{D3}}$ is given by \eqref{eqn:stress energy D3},
and the ellipses denote terms suppressed by powers of $e^{A}$.

It follows that the ten-dimensional equation of motion \eqref{eqn:Master Final}, incorporating the total stress-energy $T_{\mu\nu}^{\langle\lambda\lambda\rangle}+p\,T_{\mu\nu}^{\overbar{D3}}$ in \eqref{thetotTapproxyes},
requires the Einstein-frame scalar curvature $\mathcal{R}_4[g]$ to take exactly the value computed in the de Sitter vacuum of the four-dimensional
theory in \cite{KKLT}.
In other words, the precise quantitative match between ten-dimensional and four-dimensional computations that we established for the $\mathcal{N}=1$ supersymmetric theory in \S\ref{sec:SUSY} continues to hold in the presence of anti-D3-branes.

\section{Conclusions}  \label{sec:conclusions}

We have derived the four-dimensional scalar potential in the de Sitter and anti-de Sitter constructions of \cite{KKLT} directly from type IIB string theory in ten dimensions, supplemented with the expectation value $\langle \lambda\lambda\rangle$ of the D7-brane gaugino bilinear.

We first computed the two-gaugino and four-gaugino couplings on D7-branes, by dimensionally reducing and T-dualizing the ten-dimensional type I supergravity action.  From these terms we computed the ten-dimensional stress-energy sourced by gaugino condensation on a stack of D7-branes, carefully accounting for the fact that the ten-dimensional solution in the presence of the condensate is a generalized complex geometry.
As a key step in this computation, we used the ten-dimensional Killing spinor equations to prove that in a supersymmetric configuration, the generalized complex geometry superpotential \eqref{eqn:flux susy} is equal to the full superpotential, i.e.~we established the relation \eqref{eqn:wgcgvev}.  Upon dimensional reduction, the ten-dimensional stress-energy of the supersymmetric configuration then gives rise to the scalar potential of the $\mathcal{N}=1$ supersymmetric theory of \cite{KKLT}, evaluated in its supersymmetric $AdS_4$ vacuum.  The match is exact, at the level of the approximations made in \cite{KKLT}.  Furthermore, provided that \eqref{eqn:wgcgvev} continues to hold off-shell --- which we find plausible but have not established here --- we recovered the complete scalar potential of the four-dimensional theory, even away from the supersymmetric minimum of the potential for the K\"ahler modulus.

To combine the stress-energy of the gaugino condensate with that of anti-D3-branes at the tip of a Klebanov-Strassler throat, we examined the Kaluza-Klein spectrum of $T^{1,1}$, and found the operators of the dual field theory that mediate the leading interactions between a condensate in the ultraviolet and anti-D3-branes in the infrared.
We found that all such couplings via Kaluza-Klein excitations are suppressed by powers of the warp factor compared to the probe anti-D3-brane action.  This left the interaction via the breathing mode, as in \cite{KKLT}, as the only important one.  We thus concluded that the ten-dimensional stress-energy of
the gaugino condensate and the anti-D3-branes together lead to the scalar potential of the non-supersymmetric theory of \cite{KKLT}.  The match is again exact, even away from the de Sitter minimum, in the same sense as above.

This work has not altered the evidence, which we judge to be robust \cite{sk:fluxes}, for the existence in string theory of the separate components of the scenario \cite{KKLT}, namely a small classical flux superpotential, a gaugino condensate on a stack of D7-branes, and a metastable configuration of anti-D3-branes in a Klebanov-Strassler throat.
Instead, we showed that \emph{provided} these components exist in an explicit string compactification,
their effects can be computed either in ten dimensions or in the four-dimensional effective theory, with perfect agreement.

Progress in understanding the physics of de Sitter space in string theory continues.
Our findings may aid in pursuing de Sitter solutions in ten dimensions.

\section*{Acknowledgments}
We thank Naomi Gendler, Arthur Hebecker, Ben Heidenreich, Jakob Moritz, Gary Shiu, Pablo Soler, Irene Valenzuela, Thomas Van Riet, Alexander Westphal, and Edward Witten for discussions.  We are grateful to Iosif Bena, Mariana Gra\~{n}a, Nicolas Kovensky, Luca Martucci, and Ander Retolaza for pointing out errors in the explanations of generalized complex geometry in the first version of this work, to Jakob Moritz for making valuable suggestions about the material in \S\ref{wgravapp}, and to Anatoly Dymarsky and Luca Martucci for explanations of \cite{Dymarsky:2010mf}.  We are indebted to the anonymous referee for asking insightful questions that led to very significant improvements of this work.  This research of S.K.~was supported by NSF grant PHY-1720397 and by a Simons Investigator Award.
The work of M.K.~and L.M.~was supported in part by NSF grant PHY-1719877.  Portions of this work were completed at the Aspen Center for Physics, which is supported by National Science Foundation grant PHY-1607611.  L.M.~thanks the organizers of String Phenomenology 2019 and Strings 2019 for providing opportunities to present this work.

\begin{appendix}
\addtocontents{toc}{\protect\setcounter{tocdepth}{1}}
\section{Dimensional Reduction} \label{app:conv}

In this appendix we first obtain, in \S\ref{appa1} and \S\ref{appa2}, the couplings of D7-brane gauginos that are required for our analysis. Then, in \S\ref{wgravapp} and \S\ref{kapp} we give details of the superpotential and K\"ahler potential, respectively, in the four-dimensional theory.  Our conventions are as in \cite{Polchinski2}, augmented by $(2\pi)^2\alpha'=1$.

\subsection{D7-brane gaugino action}\label{appa1}

We first compactify type I superstring theory on $T^2$ and T-dualize to find the action on type IIB D7-branes.
As the ten-dimensional $\mathcal{N}=1$ supergravity action with a vector multiplet, including the four-gaugino action, is well known, we can determine with precision the D7-brane gaugino action including four-gaugino terms.

One minor complication is that some fields, such as the NS-NS two-form $B$, are projected out in type I superstring theory.  We will therefore first arrive at a D7-brane action containing all terms that do not involve such fields, but this will not yet be the full D7-brane action. To obtain the proper gaugino-flux coupling, one can then $SL(2,\Bbb{Z})$ covariantize the gaugino-flux coupling, following \cite{Koerber:2005qi,Dymarsky:2010mf}.

The type I supergravity action in ten-dimensional Einstein frame is \cite{Dine:1985rz,deRoo:N1sugra,Kallosh:2019oxv}
\begin{align}
S=&\frac{1}{2\kappa_{10}^2}\int \sqrt{-G}\left[\mathcal{R}_{10}-\frac{1}{2}\diff_A \phi\diff^A\phi-\frac{e^\phi}{12}\left(F_{ABC}-\frac{1}{4}e^{-\phi/2}\tr\,\bar{\chi}\Gamma_{ABC}\chi \right)^2 \right.\nonumber\\
&\qquad\qquad\qquad\left.-\frac{e^{\phi/2}}{16\sqrt{2}\pi^2} \tr\, F_{AB}F^{AB}-\tr\,\bar{\chi}\Gamma^AD_A\chi \right],
\end{align}
where $\chi$ is a 32-component Majorana-Weyl spinor. Traces here are taken in the vector representation of $SO(32)$. In order to simplify T-duality, we first rescale to string frame, using $G\mapsto e^{-\phi/2}G$.
Compactifying on a $T^2$ with volume $1/2t$, we find
\begin{equation}
S = \frac{1/2t}{2\kappa_{10}^2} \int \sqrt{-G_8} \Bigl[ e^{-2\phi}\mathcal{R}_{8} + \ldots \Bigr] .
\end{equation}
Next, we T-dualize; since we are in type I string theory, this replaces the $T^2$ by a $T^2/\Bbb{Z}_2$ with volume $t$, and re-defines $e^{-2\phi}\mapsto 2t^2 e^{-2\phi}$, yielding the eight-dimensional action
\begin{equation}
S = \frac{t}{2\kappa_{10}^2} \int \sqrt{-G_8} \Bigl[ e^{-2\phi} \mathcal{R}_8 + \ldots \Bigr]\,.
\end{equation}
Finally, we rescale back to ten-dimensional Einstein frame, using $G\mapsto e^{\phi/2} G$.

This procedure yields the new Yang-Mills term
\begin{equation}\label{theym}
\frac{1}{2\kappa_{10}^2}\cdot \frac{1}{8\pi^2} \int \sqrt{-G_8} \left[-\frac{1}{4}\tr\, F_{ab}F^{ab} \right] .
\end{equation}
Here $a,b \in \{0,\ldots,7\}$, and we will later use $i,j\in \{8,9\}$. The action \eqref{theym} is consistent with the Einstein-frame D7-brane Dirac-Born-Infeld action
\begin{equation}
-\frac{\mu_7}{2} \int \tr\left\{e^{\phi} \sqrt{-\det(G+e^{-\phi/2}F/2\pi)}\right\}.
\end{equation}
The factor of $1/2$ is due to the fact that the gauge group is $SO(2n)$; Higgsing to $U(n)$ by moving away from an O7-plane eliminates this factor (cf.~\cite{Polchinski2}).

It is now convenient to take the $T^2$ in the type I frame to have the coordinate range $[0,1]^2$, and to use the same coordinates for the double cover of the type IIB $T^2/\Bbb{Z}_2$. For simplicity, we also take the type I torus to be a square torus with string frame metric $g_{ij} = \frac{1}{2t}\delta_{ij}$. This means that the string frame metric transforms via $G_2 \mapsto G_2/(2t)^2$.

We can now study the fermionic action of the D7-brane in Einstein frame. Since we are interested in studying D7-branes on a holomorphic divisor, we will eventually take $\tr\,\bar\chi \Gamma_{ABC}\chi$ to be a linear combination of the (pullback of the) holomorphic three-form and its complex conjugate, and we can therefore retain only functions of $\tr\,\bar\chi \Gamma_{abi}\chi$.
Other contractions do not contribute to the terms of interest.

With that restriction, after T-dualizing we find the string-frame D7-brane gaugino action
 \begin{equation}
S_{\mathrm{ferm}} = \mu_7\int\sqrt{-G_8} \left[- e^{-\phi} \tr\, \bar\chi \Gamma^a D_a\chi + \frac{1}{8} F_{abi} \tr\, \bar\chi \Gamma^{abi}\chi - \frac{1}{64t}\bigl(\mathrm{tr}\,\bar\chi\Gamma_{abi}\chi\bigr)^2 \right]\,,
\end{equation} and the corresponding
Einstein-frame D7-brane gaugino action
\begin{equation}
S_{\mathrm{ferm}}= \mu_7 \int\sqrt{-G_8} \left[- \tr\,\bar\chi \Gamma^a D_a\chi + \frac{1}{8} e^{\phi/2} F_{abi} \tr\, \bar\chi\Gamma^{abi}\chi - \frac{1}{64 t_E} \bigl(\mathrm{tr}\, \bar\chi\Gamma_{abi}\chi\bigr)^2 \right]\,, \label{eqn:D7-brane gaugino action1}
\end{equation} where we have introduced the Einstein frame volume $t_E := te^{-\phi/2}$.

We remark in passing that the gaugino quartic term has a prefactor $1/t$ that depends not just on fields localized to the D7-brane, but also on the volume $t$ of the space transverse to the D7-branes.  One could wonder how such a coupling arises in a local action (we thank the referee for comments on this point).  To understand this, we consider the T-dual configuration of a stack of D3-branes transverse to $T^6/\Bbb{Z}_2$.
The four-dimensional supergravity resulting upon compactification contains a quartic gaugino term whose coefficient is  proportional to $M_{\rm{pl}}^{-2}$, or in ten-dimensional terms is proportional to $1/{\rm{Vol}}_{T^6/\Bbb{Z}_2}$.
Upon T-dualizing four times, $1/\text{Vol}_{T^6/\Bbb{Z}_2}$ is replaced by $1/t$.  Thus, the
D7-brane gaugino quartic term is a Planck-suppressed interaction that is T-dual to a local coupling required by four-dimensional $\mathcal{N}=1$ supergravity coupled to vector multiplets.

Leaving implicit henceforth that $ABC$ is a permutation of $abi$, the D7-brane gaugino action can be written in the more symmetric form
\begin{align}
S_{\mathrm{ferm}}= -\mu_7\int \sqrt{-G_8} \left[\tr\,\bar{\chi}\Gamma^AD_A\chi-\frac{e^{\phi/2}}{24} F_{ABC}\tr\,\bar{\chi}\Gamma^{ABC}\chi\right.
\left.+\frac{1}{192t_E}\Bigl(\tr\,\bar{\chi}\Gamma^{ABC}\chi\Bigr)^2\right]\,.\label{eqn:D7-brane gaugino action einstein}
\end{align}
In \eqref{eqn:D7-brane gaugino action einstein} we have obtained the part of the action that survived the type I projections.  The full D7-brane action is then given by $SL(2,\Bbb{Z})$-covariantizing.
As doing so would involve studying the transformation properties of the D7-brane fields under $SL(2,\Bbb{Z})$, which would take us too far from our main aims, and the full set of two-gaugino terms in the $\kappa$-symmetric D7-brane action was found in \cite{Koerber:2005qi,Dymarsky:2010mf}, we simply $SL(2,\Bbb{Z})$-covariantize the action by including the missing terms found by \cite{Koerber:2005qi,Dymarsky:2010mf}, leading to
\begin{align}
S_{\mathrm{ferm}}=\mu_7 \int \sqrt{-G_8}&\left[-\,\tr\,\bar{\chi}\Gamma^AD_A\chi-\frac{e^{\phi/2}}{24}\tr\,\bar{\chi}\Gamma^{ABC}\Bigl(i\tilde{F}_{ABC}\sigma_1-ie^{-\phi}H_{ABC}\sigma_3\Bigr)\chi\right.\nonumber\\
-&\left.\frac{1}{192 t_E}\Bigl(\tr\,\bar{\chi}\Gamma_{ABC}\sigma_1\chi\Bigr)^2 \right]\,,\label{eqn:final D7-brane gaugino action1}
\end{align} where the $\sigma$ matrix notation will be explained below.

\subsection{Reduction of the D7-brane action on a divisor}\label{appa2}

Equipped with the gaugino action \eqref{eqn:final D7-brane gaugino action1}, we now consider wrapping D7-branes on a divisor $D$ in an orientifold $M$ of a Calabi-Yau threefold.
We assume that there is a single K\"ahler modulus $T$, with the K\"ahler form written as
\begin{equation}
J=t\omega\,,
\end{equation}
and the volume
\begin{equation}\label{thevis}
\mathcal{V}e^{6u} =\frac{1}{3!} t^3\,,
\end{equation}
where we have normalized $\omega\in H_+^2(M,\Bbb{Z})$ such that $\int_M\omega\wedge\omega\wedge\omega=1$, and we have normalized $e^{-4A}$ such that $\int_M e^{-4A} \omega\wedge\omega\wedge \omega=1$.
We take the volume of $D$ to be
\begin{equation}
\int_D\sqrt{g} e^{-4A+4u}=\text{Re}(T)=t^2/2\,,
\end{equation}
while the volume of the curve dual to $D$ is $t$, and corresponds to $t_E$ in \eqref{eqn:final D7-brane gaugino action1}.  The divisor $D$ is assumed to be rigid, and so the D7-branes will not explore the transverse space, and therefore the geometry of the latter is unimportant.  However, for later use we record that the volume of the transverse space is
\begin{equation}
\mathcal{V}_\perp e^{2u}=\frac{1}{3} t\,.
\end{equation}

We note that wrapping on $D$ topologically twists the D-brane worldvolume theory, so that scalars become sections of the normal bundle $N$ of $D$ and fermions become spinors on the total space of this normal bundle \cite{vafa:k3inst}. For notational convenience, we implement the topological twist via a background $U(1)$ R-symmetry gauge field, rather than by re-defining the local Lorentz group. Since, locally, the Calabi-Yau manifold looks like the total space of the normal bundle, there is no topological obstruction to relating these fermions to the covariantly constant spinor on the Calabi-Yau.

\subsubsection{Internal spinors}

As our ansatz for the geometry of the internal space $M$, we take $M$ to have an $SU(2)$ structure.
This can be encoded in terms of two globally-defined orthonormal spinors, $\eta_+$ and $\chi_+$, and an invariant one-form $v_a dy^a$, that are related by
\begin{equation}
\chi_+=\frac{1}{2}v^a\gamma_a\eta_+^{*}\,,
\end{equation}
where $|v|^2=2$.
Using $\chi_+$ and $\eta_+$ one can construct invariant forms with the components
\begin{equation}
v^a=\eta_+^T\gamma^a\chi_+\,,\, J_2^{mn}=i\eta_+^\dagger \gamma^{mn}\eta_+-i\chi_+^\dagger\gamma^{mn}\chi_+\,,\, \Omega_2^{mn}=-i\chi_+^\dagger \gamma^{mn}\eta_+\,,
\end{equation}
\begin{equation}
J^{mn}=i\eta_+^\dagger \gamma^{mn}\eta_+\,,\, \Omega^{mnp}=-i\eta_+^T\gamma^{mnp}\eta_+\,.
\end{equation}
The invariant forms satisfy
\begin{equation}
J_2\wedge\Omega_2=\Omega_2\wedge\Omega_2=0\,,\, v_a \Omega_2^{ab}=v_a J_2^{ab}=0\,,\,J_2\wedge J_2=\frac{1}{2}\Omega\wedge\overline{\Omega}_2\,,
\end{equation}
\begin{equation}
J=J_2+\frac{i}{2}v\wedge\overline{v}\,,\, \Omega=\Omega_2\wedge v\,.
\end{equation}
We now construct the linear combinations
\begin{equation}
\eta_1:=ie^{A/2+i\vartheta/2} \left(\cos\frac{\varphi}{2}\eta_++\sin\frac{\varphi}{2}\chi_+ \right)\,,\label{eq:newConv1}
\end{equation}
\begin{equation}
\eta_2:=e^{A/2-i\vartheta/2}\left(\cos\frac{\varphi}{2}\eta_+-\sin\frac{\varphi}{2}\chi_+\right)\,, \label{eq:newConv2}
\end{equation}
which are normalized as
\begin{equation}
\eta_1^\dagger\eta_1=\eta_2^\dagger\eta_2=e^{A}.
\end{equation}
The parameters $\varphi$ and $\vartheta$ represent the angles between $\eta_1$ and $\eta_2$: from
\eqref{eq:newConv1} and \eqref{eq:newConv2} one has
\begin{equation}
\eta_2^\dagger\eta_1=i e^{i\vartheta+A} \cos{\varphi}\,.
\end{equation}
The spinors $\eta_1$ and $\eta_2$ can be repackaged into a pair of bispinors:
\begin{equation}
\Phi_1:=-8ie^{-A}\eta_1\otimes\eta_2^\dagger\,,
\end{equation}
\begin{equation}
\Phi_2:=-8ie^{-A}\eta_1\otimes\eta_2^T\,.
\end{equation}
Using the Clifford map, $\Phi_1$ and $\Phi_2$ are polyforms: specifically, they can be written in terms of invariant forms as
\begin{equation}
\Phi_1=e^{i\vartheta}e^{\frac{1}{2}v\wedge\overline{v}}\left[\cos\varphi \left(1-\frac{1}{2}J_2\wedge J_2 \right) -iJ_2+\sin\varphi \im\Omega_2\right]\,,
\end{equation}
\begin{equation}
\Phi_2=v\wedge \left[i\re\Omega_2-\cos\varphi\im\Omega_2+\sin\varphi\left( 1-\frac{1}{2}J_2\wedge J_2\right)\right]\,.
\end{equation}

The ansatz we have just described corresponds to a generic SU(2) structure.  If $M$ is a Calabi-Yau orientifold then in fact $e^{i\vartheta}=1$ and $\varphi=0$.  However, once gaugino condensation is incorporated and $M$ becomes a generalized complex geometry, $\varphi$ will vary non-trivially along $M$; the SU(2) structure is then said to be dynamic.

We now expand to first order in the small quantity $\langle\lambda\lambda\rangle$, using the fact that $\varphi=\mathcal{O}(\langle\lambda\lambda\rangle)$.
We find
\begin{equation}
\Phi_1=e^{-iJ}\Bigl(1+\varphi\im\Omega_2\Bigr)+\mathcal{O}\bigl(\langle\lambda\lambda\rangle^2\bigr)\,,\label{eqn:phi1 approx}
\end{equation}
\begin{equation}
\Phi_2=i\Omega+\varphi~ v\wedge\left( 1-\frac{1}{2}J_2\wedge J_2\right)+\mathcal{O}\bigl(\langle\lambda\lambda\rangle^2\bigr)\,,\label{eqn:phi2 approx}
\end{equation} while the two-form component of $\mathfrak{t}$ is
\begin{equation}
\mathfrak{t}=+e^{-\phi/2-2A}\varphi \im\Omega_2+\mathcal{O}\bigl(\langle\lambda\lambda\rangle^2\bigr)\,.\label{eqn:mathfrakt}
\end{equation}
On neglecting the terms of order $\langle\lambda\lambda\rangle^2$, $\Phi_1$ and $\Phi_2$ reduce to the $\beta$-deformed pure spinors found in \cite{Dymarsky:2010mf}.

\subsubsection{Ten-dimensional spinor ansatz}

Equipped with the six-dimensional spinors $\eta_1$ and $\eta_2$, we can now give our ansatz for the ten-dimensional spinors.
The $SL(2,\Bbb{Z})$-covariant $\kappa$-symmetric D7-brane action is usefully written in a redundant notation, involving two copies of the ten-dimensional fermion \cite{Lust:2008zd,Heidenreich:2010ad}, which we now adopt.
We consider a doublet $\chi=(\chi_1,\chi_2)$ of 32-component ten-dimensional Majorana-Weyl spinors, and decompose these spinors under ${\rm Spin}(10)\to {\rm Spin}(4)\times {\rm Spin}(6)$.
The ten-dimensional gamma matrices decompose as
\begin{equation}
\Gamma^\mu=e^{-A+3u}\gamma^\mu\otimes 1\,, \qquad \Gamma^i=e^{A-u}\gamma^5\otimes \gamma^i\,.
\end{equation}
For gamma matrices and spinor manipulations, we use the conventions of \cite{VanProeyen:1999ni},
\begin{equation}
\gamma_0=\left(\begin{array}{cc} 0 & i \\ i & 0 \end{array}\right), \  \gamma_i=\left(\begin{array}{cc}0 & -i\sigma_i \\ i\sigma_i &0 \end{array} \right), \  \gamma_5=\left(\begin{array}{cc} 1 & 0 \\ 0& -1 \end{array}\right),\  \mathcal{C}=\left(\begin{array}{cc} \epsilon & 0 \\ 0 & -\epsilon \end{array}\right),\  \epsilon=\left(\begin{array}{cc} 0 &1 \\ -1 & 0 \end{array}\right).
\end{equation}
Under this decomposition, a ten-dimensional Weyl spinor decomposes as $\mathbf{16}_+ \mapsto (\mathbf{2}_+\otimes \mathbf{4}_+)\oplus (\mathbf{2}_-\otimes \mathbf{4}_-)$, where subscripts denote chirality. We can thus write the ten-dimensional Majorana-Weyl spinors as
\begin{equation}
\chi_{1}=\frac{1}{4\pi} e^{-2A+9u/2}\,\lambda_D\otimes \eta_1+c.c.\label{eqn:spinor decomposition 1}
\end{equation}
and
\begin{equation}
\chi_2=-\frac{1}{4\pi} e^{-2A+9u/2}\,\lambda_D\otimes \eta_2+c.c.\label{eqn:spinor decomposition 2}
\end{equation}
where $c.c.$ refers to charge conjugation, and $\lambda_D$ is the embedding of a four-dimensional Weyl spinor $\lambda$ into a Dirac spinor via
\begin{equation}
\lambda_D=\left(\begin{array}{c}
0\\
\bar{\lambda}^{\dot{\alpha}}
\end{array} \right)\,.
\end{equation}

\subsubsection{Decomposition of D7-brane action}\label{decompapp}

We can now expand the D7-brane action \eqref{eqn:final D7-brane gaugino action1} in terms of the spinors in \eqref{eqn:spinor decomposition 1} and \eqref{eqn:spinor decomposition 2}.  We will henceforth leave traces implicit, writing
\begin{equation}
\tr\, \chi\chi=\frac{1}{2} \chi^a\chi^a=\frac{1}{2}\chi\chi\,,
\end{equation}
with the normalization
\begin{equation}
\tr\, T^a T^b=\frac{1}{2}\delta^{ab}
\end{equation}
for Lie algebra generators.  We likewise leave implicit pullbacks to the divisor $D$.

The gaugino kinetic term can be decomposed as
\begin{equation}
S_{\mathrm{kin}}=-\mu_7 \int_{X\times D} \sqrt{-G}\,\tr\,  \bar{\chi}\Gamma^AD_A\chi = \int_{X\times D}\sqrt{-G}\,\Bigl(\mathcal{L}_{\mathrm{kin},X}+\mathcal{L}_{\mathrm{kin},D}\Bigr)\,,
\end{equation}
with
\begin{align}
\int_{X\times D}\sqrt{-G}\,\mathcal{L}_{\mathrm{kin},X} =& -2\pi\int_{X\times D}\sqrt{-G}\,\tr\, \bar{\chi}\Gamma^\mu D_\mu\chi\\
=& -\frac{i}{4\pi}\int_{X\times D}\sqrt{-g}  e^{-4A+4u} \bar{\lambda}\bar{\sigma}^\mu D_\mu\lambda  \\
=&-\frac{i}{4\pi}\int_X \sqrt{-g}\,\text{Re}(T) \bar{\lambda}\bar{\sigma}^\mu D_\mu\lambda\,,
\end{align}
and
\begin{align}
\mathcal{L}_{\mathrm{kin},D}=&-2\pi \,\tr\,\bar{\chi}\Gamma^aD_a\chi\\
=&~\frac{1}{16\pi}\bar{\lambda}^c_D \lambda_D\left(\eta_1^TD_a (e^{-3A}\gamma^a\eta_1)+\eta_2^TD_a(e^{-3A}\gamma^a\eta_2)\right)+c.c.\\
=&-\frac{1}{16\pi}  \bar{\lambda}\bar{\lambda}\left(\eta_1^TD_a (e^{-3A}\gamma^a\eta_1)+\eta_2^TD_a(e^{-3A}\gamma^a\eta_2)\right)+c.c.\\
=&-\frac{1}{128\pi} \bar{\lambda}\bar{\lambda}\left((e^{-A}\eta_1^T\gamma_{123}\eta_2)\eta_2^\dagger \gamma^{123}D_a(e^{-3A}\gamma^a\eta_1)+(\eta_1\leftrightarrow\eta_2)\right)+c.c.\\
=&\,\frac{i}{32\pi} e^{-2u+\phi/2}\bar{\lambda}\bar{\lambda}\,i \mathrm{d}_2\mathfrak{t}\cdot \Omega+c.c.\,\label{eqn:GCG correction}
\end{align}
where we have defined $\mathrm{d}_2\mathfrak{t}=\partial_a \mathfrak{t}dz^a+\partial_{\bar{a}}\mathfrak{t}d\bar{z}^{\bar{a}}.$
Here $a \in \{1,2\}$, where $z_1$ and $z_2$ are complex coordinates along the D7-brane divisor $D$, and we stress that $\mathfrak{t}$ in \eqref{eqn:GCG correction} must be understood as the pullback onto $D$ of the form $\mathfrak{t}$ defined in $M$.

In \eqref{eqn:GCG correction} we have omitted terms that are higher order in $\langle\lambda\lambda\rangle$, in particular the terms of order $\langle\lambda\lambda\rangle^2$ in
\eqref{eqn:phi1 approx}, \eqref{eqn:phi2 approx}, and \eqref{eqn:mathfrakt}.
We make the same approximation in the computations below.

For the gaugino-flux couplings, we find
\begin{align}
\mathcal{L}_{F\lambda\lambda}=&~i\frac{e^{\phi/2-2u-A}}{384\pi} \left( \bar{\lambda}_D\otimes \eta_1^\dagger +\bar{\lambda}^c_D\otimes \eta_1^T\right)\tilde{F}_{ABC}\gamma^{ABC}\gamma^5\left(\lambda_D\otimes\eta_2+\lambda^c_D\otimes \eta_2^*\right)+(\eta_1\leftrightarrow\eta_2)\label{threeformonly}\\
=&-i\frac{e^{\phi/2-2u-A}}{384\pi}  (\bar{\lambda}^c_D \lambda_D\eta_1^T\gamma^{ABC}\eta_2+c.c.)\tilde{F}_{ABC}+(\eta_1\leftrightarrow \eta_2)\\
=&\frac{ie^{\phi/2-2u}}{32\pi}  \bar{\lambda}\bar{\lambda}\, \tilde{F}\cdot \Omega+c.c.\label{eqn:F-gaugino}
\end{align}
\begin{align}
\mathcal{L}_{H\lambda\lambda}=&i\frac{e^{-\phi/2-2u-A}}{384\pi}  \left( \bar{\lambda}_D\otimes \eta_1^\dagger +\bar{\lambda}^c_D\otimes\eta_1^T\right)H_{ABC}\gamma^{ABC}\gamma^5\left(\lambda_D\otimes\eta_1+\lambda^c_D\otimes\eta_1^*\right)+(\eta_1\leftrightarrow\eta_2),\\
=&-i\frac{ e^{-\phi/2-2u-A}}{384\pi}\Bigl(\bar{\lambda}_D^c \lambda_D\eta_1^T\gamma^{ABC}\eta_1+c.c.\Bigr)H_{ABC}+(\eta_1\leftrightarrow\eta_2)\\
=&\frac{e^{-\phi/2-2u}}{32\pi}  \bar{\lambda}\bar{\lambda}\, H\cdot \Omega+c.c.\label{eqn:H-gaugino}
\end{align}

We should point out that in \eqref{eqn:F-gaugino} and \eqref{eqn:H-gaugino} only the three-form fluxes appear, in contrast to the democratic formulation of generalized complex geometry in which three-forms and seven-forms enter on equal footing.  One might then worry that the deformation of the background due to gaugino condensation could introduce corrections to the action of the Hodge star on internal forms, and in turn to the effective action.  (We thank the referee for raising this issue.)  However, from \eqref{eqn:phi2 approx} one finds that the three-form component of $\Phi_2$ is not corrected at order $\mathcal{O}(\langle\lambda\lambda\rangle)$.  Thus, the Hodge star acting on internal three-forms is not corrected at order $\mathcal{O}(\langle\lambda\lambda\rangle)$, and the resulting corrections to the effective action are smaller than order $\mathcal{O}(\langle\lambda\lambda\rangle^2)$, and can therefore be neglected in our analysis.

Combining \eqref{eqn:F-gaugino} and \eqref{eqn:H-gaugino}, we obtain the coupling
\begin{equation}
S_{G\lambda\lambda} = \frac{i}{32\pi}\int_{X\times D}\sqrt{-g} e^{-2u+\phi/2}\bar{\lambda}\bar{\lambda}\,G\cdot \Omega+c.c.\label{eqn:CIU}
\end{equation}
Thus, combining \eqref{eqn:CIU} and \eqref{eqn:GCG correction}, the total gaugino-flux coupling is
\begin{equation}
S_{\mathfrak{G}\lambda\lambda} = \frac{i}{32\pi}\int_{X\times D}\sqrt{-g} e^{-2u+\phi/2}\bar{\lambda}\bar{\lambda}\,\mathfrak{G}^{[2]}\cdot\Omega+c.c.\label{eqn:the flux-gaugino coupling}
\end{equation}
The result \eqref{eqn:the flux-gaugino coupling} precisely agrees with that of \cite{Dymarsky:2010mf} once one accounts for the difference in normalization of the gaugino kinetic term there and here.

Similarly, we find the four-gaugino couplings
\begin{align}
\mathcal{L}_{\lambda\lambda\lambda\lambda}=&-\frac{e^{10u-6A}}{3\cdot 2^{15}\pi^3 \,t}\Bigl[(\bar{\lambda}_D \otimes \eta_1^\dagger +c.c.) \gamma^{abc}(\lambda_D\otimes\eta_2+c.c.)+(\eta_1\leftrightarrow\eta_2)\Bigr]^2\\
=&- \frac{e^{10u-6A}}{3\cdot 2^{15}\pi^3\,t}\Bigl[ \bar{\lambda}^c_D\lambda_D \eta_1^T\gamma^{abc}\eta_2+c.c.+(\eta_1\leftrightarrow \eta_2)\Bigr]^2\\
=&- \frac{e^{10u-4A}}{3\cdot 2^{15}\pi^3\,t}\Bigl[2i \bar{\lambda}\bar{\lambda}\Omega^{abc}-2i\lambda\lambda\overline{\Omega}^{abc}\Bigr]^2\\
=&-\frac{\nu e^{8u-4A}}{6144\pi^3} \Omega\cdot\overline{\Omega}\, \lambda\lambda\bar{\lambda}\bar{\lambda}\,, \label{finalfourg}
\end{align} where $\nu$ was defined below \eqref{eqn:fourgaugino}.

We have thus obtained the Lagrangian density for D7-brane gauginos, up to and including $|\lambda\lambda|^2$ terms:
\begin{equation}
\boxed{\vphantom{\Biggl(\Biggr)}\mathcal{L}_{\mathrm{gaugino}}=-\frac{i}{4\pi} e^{-4A+4u} \bar{\lambda}\bar{\sigma}^\mu\diff_\mu\lambda+\frac{i}{32\pi} e^{-2u+\phi/2}\,\mathfrak{G}^{[2]}\cdot \Omega\, \bar{\lambda}\bar{\lambda}+c.c.-\frac{\nu}{6144\pi^3}e^{8u-4A} \Omega\cdot\overline{\Omega} \left|\lambda\lambda \right|^2}
\end{equation}

\subsection{Killing spinor equations and the superpotential}\label{wgravapp}

The overall goal of this work has been to determine whether the ten-dimensional field configuration that results when gaugino condensation is taken as a  source in the ten-dimensional equations of motion ultimately leads to a four-dimensional scalar potential that exactly matches that computed in the four-dimensional supergravity theory of \cite{KKLT}.  In order to perform this comparison, we must \emph{translate} the data of the ten-dimensional fields into four-dimensional expressions.  Specifically, we need to express the gaugino-flux coupling \eqref{eqn:ciugcg} in terms of the superpotential $W$, by  relating the generalized flux $\mathfrak{G}$ to $W$.  In this section we carefully explain the correspondence between ten-dimensional and four-dimensional data.

\subsubsection{Outline}\label{sssappsumm}

As a guide through the computations ahead, we first outline our logic.  First, building on \cite{Koerber:2007xk,Bena:2019mte}, we write down the ten-dimensional Killing spinor equations whose solutions are supersymmetric configurations.  The classical Killing spinor equations are well-known, and the difficulty lies in modifying them to account for the effect of gaugino condensation.  To determine the correct modification, we demand the following consistency conditions:
\begin{enumerate}
  \item The three-form fluxes $G_{0,3}$, $G_{3,0}$, and $G_{1,2}$ obtained from the Killing spinor equations must be compatible with the solution of the Bianchi identities.\label{cond1}
  \item The IASD three-form flux $G_{3,0}$ obtained from the Killing spinor equations must vanish in the vacuum configuration: nonvanishing $G_{3,0}$ would give mass to the gaugino on a probe D3-brane, and so is incompatible with supersymmetry.\footnote{We thank Jakob Moritz for suggesting this condition.}\label{cond2}
\end{enumerate}
We write down a very general modification of the classical Killing spinor equations, involving three a priori independent terms proportional to $\langle\lambda\lambda\rangle$, with initially undetermined coefficients, and show that the above conditions uniquely determine the values of all three coefficients.  As we explain in detail below, the resulting Killing spinor equations \eqref{killing1final}-\eqref{killing3final} are \emph{not} exactly those of \cite{Koerber:2007xk}, which contain only a single term proportional to $\langle\lambda\lambda\rangle$.  We believe that the consistency conditions above are compulsory, independent of any attempt to argue for or against a ten-dimensional description of the de Sitter vacua of \cite{KKLT}, and so we claim that our modified Killing spinor equations \eqref{killing1final}-\eqref{killing3final} are the correct ones in this setting.

We then turn to the superpotential $W_{\mathrm{GCG}}$ \eqref{eqn:superpotential}
that has been argued to govern a general type IIB string compactification on a generalized complex geometry \cite{Grana:2005ny,Benmachiche:2006df,Micu:2007rd,Koerber:2007xk,Lust:2008zd}.
Computing the expectation value $\langle W_{\mathrm{GCG}}\rangle$ on the solution of the Killing spinor equations, we
find that $\langle W_{\mathrm{GCG}}\rangle$ equals the full superpotential $W$ when the Killing spinor equations are the corrected ones that we justified above, but that $\langle W_{\mathrm{GCG}}\rangle \neq W$ when the Killing spinor equations are those given in \cite{Koerber:2007xk,Bena:2019mte}.  Correspondingly, we demonstrate that \emph{using the corrected Killing spinor equations \eqref{killing1final}-\eqref{killing3final} we exactly recover the scalar potential of \cite{KKLT} from ten dimensions.}

\subsubsection{Gaugino condensation and the Killing spinor equations}

We begin with a rather general form of the Killing spinor equations,
\begin{align}
d_H \Bigl(e^{(\phi/4-A)\hat{p}}e^{3A-\phi/4}\Phi_2\Bigr) =~&2i \mu e^{(\phi/4-A)\hat{p}}e^{2A-\phi/2}\im\Phi_1+2\alpha \langle S\rangle \delta^{(2)},\label{killing1}\\
d_H\Bigl(e^{(\phi/4-A)\hat{p}}e^{2A-\phi/2}\im\Phi_1\Bigr) =~&0,\label{killing2}\\
d_H\Bigl(e^{(\phi/4-A)\hat{p}}e^{4A}\re\Phi_1\Bigr) =~&3 e^{(\phi/4-A)\hat{p}}e^{3A-\phi/4}\re(\overline{\mu}\Phi_2)+e^{(2A-\phi/2)(3-\hat{p})} e^{4A+\phi}\tilde{F}\nonumber\\  &+\frac{e^{\phi/2}}{2}\re(\langle S\rangle \overline{\Omega})\left(\beta \delta^{(0)}+\frac{\xi}{\mathcal{V}_\perp}\right)\,.\label{killing3}
\end{align}
We define $d_H:= d-H\wedge$ and $\tilde{F}=(-1)^{\hat{p}(\hat{p}-1)/2}\star_6 F.$ We have written \eqref{killing1}-\eqref{killing3} in Einstein frame, and with the notational simplification $\langle S\rangle \equiv \langle \lambda\lambda\rangle/32\pi^2$.  The parameter $\mu = -i e^{\phi/2} \kappa_4^2 W$ is determined by the full superpotential $W$,\footnote{Throughout this work, $W$ always denotes the full superpotential, as opposed to a single term in the superpotential, such as the flux superpotential term $W_{\mathrm{flux}}$.} and so is related to the cosmological constant $\Lambda$ at the supersymmetric minimum by $e^{\kappa_4^2 K/2} e^{-\phi/2}|\mu| = \sqrt{-\Lambda/3}$.

In the Killing spinor equations given in \cite{Koerber:2007xk,Bena:2019mte}, the constants $\beta$ and $\xi$ are zero, and $\alpha=1$.
We will demonstrate below that consistency actually requires $\alpha=1$, $\beta=2$, and $\xi=0$.\footnote{We could also have added a nonsingular term $2\gamma \langle S\rangle/\mathcal{V}_\perp$ to the right-hand side of \eqref{killing1}, but from the analysis below it will be easily seen that in fact $\gamma$ must vanish.  To reduce the complexity of the expressions that follow, we set $\gamma=0$ at the outset.}

\subsubsection{Fluxes and the Bianchi identities}

We now compute various fields from the Killing spinor equations. To obtain the three-form flux, we compute $\Bigl\langle \eqref{killing3},e^{(\phi/4-A)\hat{p}} e^{3A-\phi/4}\Phi_2\Bigr\rangle.$ We first examine the left-hand side of \eqref{killing3} and use \eqref{killing1} to obtain
\begin{equation}
 \Bigl\langle d_H (e^{(\phi/4-A)\hat{p}} e^{4A}\re\Phi_1),e^{(\phi/4-A)\hat{p}}e^{3A-\phi/4} \Phi_2\Bigr\rangle= \mu e^{\phi}\langle \overline{\Phi}_2,\Phi_2\rangle +i\frac{\alpha e^{\phi}}{4} \langle S\rangle \langle \overline{\Phi}_2,\Phi_2\rangle\delta^{(0)}\,,
\end{equation}
where $\langle,\rangle$ denotes the Mukai pairing, and we have used the relations $\langle\re\Phi_1,\im\Phi_1\rangle=i\langle\Phi_1,\overline{\Phi}_1\rangle/2=-i\langle\overline{\Phi}_2,\Phi_2\rangle/2$ and $\langle\re\Phi_1,\delta^{(2)}\rangle=\frac{i}{8}\langle \overline{\Phi}_2,\Phi_2\rangle \delta^{(0)}.$ We have taken the normalization
\begin{equation}\label{mukainorm}
  \langle \Phi_1,\overline{\Phi}_1\rangle =\langle \Phi_2, \overline{\Phi}_2\rangle = 8i J^3/3!+\mathcal{O}(\lambda\lambda)\,,
\end{equation}
cf.~\eqref{eqn:phi1 approx} and \S\ref{kapp}.
Using the right-hand side of \eqref{killing3}, we find
\begin{align}
\Bigl\langle d_H(e^{(\phi/4-A)\hat{p}}e^{4A}\re\Phi_1),e^{(\phi/4-A)\hat{p}}e^{3A-\phi/4}\Phi_2\Bigr\rangle=&\frac{3}{2}\mu e^\phi\langle \overline{\Phi}_2,\Phi_2\rangle+e^{4A+3\phi/2}\langle \tilde{F},\Phi_2\rangle\nonumber\\&+\frac{1}{4} e^\phi  \langle S\rangle\langle \overline{\Omega},\Phi_2\rangle\left(\beta\delta^{(0)}+\frac{\xi}{\mathcal{V}_\perp}\right)\,.
\end{align}

We therefore compute:
\begin{equation}
\tilde{F}|_{(0,3)}=\frac{i}{2} \mu e^{-4A-\phi/2} \overline{\Omega}+\frac{1}{4} e^{-4A-\phi/2} \langle S\rangle \overline{\Omega} \left((\alpha-\beta)\delta^{(0)}-\frac{\xi}{\mathcal{V}_\perp} \right)\,,
\end{equation}
\begin{align}\label{sol:killing}
\mathfrak{G}|_{(0,3)}=&i\left.\left(\tilde{F}+e^{-4A-\phi} d_H (e^{\phi/2 +2A}\re\Phi_1^{(2)})\right)\right|_{(0,3)}\\
=&\frac{1}{2} e^{-4A-\phi/2}\mu\overline{\Omega}+\frac{i}{4} e^{-4A-\phi/2} \langle S\rangle \overline{\Omega}\left((2\alpha-\beta)\delta^{(0)}-\frac{\xi}{\mathcal{V}_\perp} \right)\,,
\end{align}
\begin{equation}\label{mfg03}
\overline{\mathfrak{G}}|_{(0,3)}=-\frac{3}{2}\mu e^{-4A-\phi/2} \overline{\Omega}-\frac{i}{4} \langle S\rangle e^{-4A-\phi/2}\overline{\Omega}\left(\beta\delta^{(0)}+\frac{\xi}{\mathcal{V}_\perp}\right)\,,
\end{equation}
\begin{equation}
id\mathfrak{t}:=i e^{-4A-\phi} d (e^{\phi/2+2A}\re\Phi_1^{(2)})\,,
\end{equation}
\begin{equation}\label{idt}
id\mathfrak{t}_{(0,3)}=+\frac{i\alpha}{2} e^{-4A-\phi/2}\langle S\rangle \overline{\Omega}\left(\delta^{(0)}-\frac{1}{\mathcal{V}_\perp}\right)\,,
\end{equation}
\begin{equation}\label{G12KSE}
G_{(1,2)}=-id\mathfrak{t}_{(1,2)}=-2i\alpha e^{-4A-\phi/2}\langle S\rangle  \partial_z^2 G_{(2)}(z;z_{D7}) v\wedge\overline{\Omega}_2 \,,
\end{equation}
\begin{equation}\label{G03KSE}
G_{(0,3)}=\frac{1}{2} e^{-4A-\phi/2} \mu\overline{\Omega} -\frac{i}{4} e^{-4A-\phi/2}\langle S\rangle\left(\beta\delta^{(0)}+\frac{\xi -2\alpha}{\mathcal{V}_\perp}\right)\,,
\end{equation}
\begin{equation}\label{G30KSE}
G_{(3,0)}=-\frac{3}{2}e^{-4A-\phi/2} \bar{\mu}\Omega -\frac{i}{4} e^{-4A-\phi/2}\langle \bar{S}\rangle\Omega \left((2\alpha-\beta) \delta^{(0)}-\frac{\xi+2\alpha}{\mathcal{V}_\perp}\right)\,.
\end{equation}

Next we find the solutions of the Bianchi identities. To simplify the problem, we will assume that $d\tau=0$.  The Bianchi identities are
\begin{equation}
d G_+= dG_-\,,
\end{equation}
and
\begin{equation}
d\Lambda= dX\,,
\end{equation}
with
\begin{equation}
\Lambda= e^{4A}\star_6 G_3-i\alpha G_3\,,
\end{equation}
and, as we shall show,
\begin{equation}\label{xisdef}
X=\frac{e^{-\phi/2}}{32\pi^2} \lambda\lambda\overline{\Omega}\delta^{(0)}\,.
\end{equation}
Let us first establish \eqref{xisdef}.
Starting from the action
\begin{equation}
S_{G\lambda\lambda}=\frac{1}{32\pi } \int_{X\times D} \sqrt{-g} e^{\phi/2} \bar{\lambda}\bar{\lambda}G\wedge \Omega+c.c.\,,
\end{equation} we compute
\begin{equation}
\frac{\partial \mathcal{L}_{G\lambda\lambda}}{\partial d C_2}= \frac{e^{\phi/2}}{32\pi} d^4 x \wedge \left(\bar{\lambda}\bar{\lambda}\Omega+\lambda\lambda\overline{\Omega}\right)\delta^{(0)}\,,
\end{equation} and
\begin{equation}
\frac{\partial \mathcal{L}_{G\lambda\lambda}}{\partial d B_2}=\frac{e^{\phi/2}}{32\pi } d^4 x\wedge \left( -\tau \bar{\lambda}\bar{\lambda} \Omega-\bar{\tau}\lambda\lambda \overline{\Omega}\right)\delta^{(0)}\,,
\end{equation} so that
\begin{equation}
\tau d\left( \frac{\partial \mathcal{L}_{G\lambda\lambda}}{\partial d C_2}\right)+d\left(\frac{\partial \mathcal{L}_{G\lambda\lambda}}{\partial d B_2}\right) = \frac{ie^{-\phi/2}}{16\pi}d^4x \wedge d\left(\lambda\lambda\overline{\Omega} \delta^{(0)}\right)\,,
\end{equation} confirming \eqref{xisdef}.

At lowest order in $\mathcal{O}(\lambda\lambda)$, $\Lambda=2 e^{4A}G_-$, and so
\begin{equation}
G_-=-\frac{e^{-4A-\phi/2}}{32\pi^2}\lambda\lambda \partial_a\partial_b G_{(2)}(z;z_{D7})g^{b\bar{b}}\overline{\Omega}_{\bar{b}\bar{c}\bar{d}}\,,
\end{equation} and
\begin{equation}
G_+=\frac{e^{-4A}}{2}X=\frac{e^{-4A-\phi/2}}{64\pi^2}\lambda\lambda\overline{\Omega}\delta^{(0)}\,,
\end{equation}
so that the \emph{singular} terms in the flux are
\begin{equation}\label{G12BI}
G_{(1,2)}=iG_{-}|_{(1,2)}=-i\frac{e^{-4A-\phi/2}}{32\pi^2}\lambda\lambda \partial_a\partial_b G_{(2)}(z;z_{D7})g^{b\bar{b}}\overline{\Omega}_{\bar{b}\bar{c}\bar{d}}\,,
\end{equation} and
\begin{equation}\label{G03BI}
G_{(0,3)}=-i G_{+}|_{(0,3)}=-i\frac{e^{-4A-\phi/2}}{64\pi^2}\lambda\lambda\overline{\Omega}\delta^{(0)}\,,
\end{equation}
whereas
\begin{equation}\label{G30BI}
G_{(3,0)}=\mathrm{nonsingular}\,.
\end{equation}

\subsubsection{Consistency conditions}

As explained above, we must enforce that the Killing spinor equations are compatible with the Bianchi identities:
\begin{enumerate}
\item The three-form flux $G_{1,2}$ obtained from the Killing spinor equations must be compatible with the solution of the Bianchi identities.
     Comparing \eqref{G12KSE} and \eqref{G12BI}, this implies that $\alpha=1.$
\item The three-form flux $G_{0,3}$ obtained from the Killing spinor equations must be compatible with the solution of the Bianchi identities.
    Comparing \eqref{G03KSE} and \eqref{G03BI}, this implies that $\beta=2$.
\item The three-form flux $G_{3,0}$ obtained from the Killing spinor equations must be compatible with the solution of the Bianchi identities.
    Comparing \eqref{G30KSE} and \eqref{G30BI}, this implies that $\beta=2\alpha$.
\end{enumerate}
We conclude that $\alpha=1$ and $\beta=2$.  The normalization $\alpha=1$ agrees with \cite{Koerber:2007xk,Bena:2019mte}.  However, $\beta=0$ in \cite{Koerber:2007xk,Bena:2019mte}, so we find that consistency with the Bianchi identities requires that we include a new term in the Killing spinor equations.

The coefficient $\xi$ has not yet been fixed, but we have another consistency condition to impose:
\begin{enumerate}
\setcounter{enumi}{3}
\item The IASD three-form flux $G_{3,0}$ obtained from the Killing spinor equations must vanish in a supersymmetric vacuum.
Using $\alpha=1$ and $\beta=2$ in \eqref{G30KSE} gives
\begin{equation}\label{G30KSE2}
e^{4A+\phi/2} G_{(3,0)}=-\frac{3}{2} \bar{\mu}\Omega +\frac{i\left(\xi+2\alpha\right)}{4\mathcal{V}_\perp}\langle \bar{S} \rangle\Omega \,.
\end{equation}
Using the relations $\mu=-i e^{\phi/2} W/(4\pi \mathcal{V})$, $\langle S\rangle =\frac{1}{2\pi }e^{\kappa_4^2K/2}\partial_T W_{\mathrm{np}}$, and $\mathcal{V}= \mathcal{V}_{\perp} \re T$, we find
\begin{equation}\label{G30KSE3}
-4\pi i\mathcal{V}_{\perp}e^{4A+\phi/2} G_{(3,0)}= \Bigl\{\bigl(\tfrac{2\alpha+\xi}{2}\bigr)\overline{\partial_T W}+\overline{K_{T}W}\Bigr\}\,\Omega\,.
\end{equation}
The D3-brane gaugino mass is\footnote{We have omitted a term proportional to $d \mathfrak{t}$ in the D3-brane gaugino mass, because $\mathfrak{t}$ only varies along the coordinates of the internal manifold, and so $d\mathfrak{t}$ has no components parallel to the D3-brane worldvolume.}
\begin{align}
m_{\lambda\lambda}\propto&\int e^{4A}G\wedge\overline{\Omega}\,\delta^{(0)}(z-z_{D3})\,.
\label{md3gaug}
\end{align}
Comparing \eqref{G30KSE3} and \eqref{md3gaug}, we see that the D3-brane gaugino mass can vanish in a supersymmetric vacuum, where $F_T=0$, only if $2\alpha+\xi=2$.  We found above that $\alpha=1$, so we conclude that $\xi=0$.
\end{enumerate}

In sum, we obtain
\begin{equation} \label{valuesofcoeffs}
\alpha=1,\qquad\beta=2,\qquad \text{and}~~\xi=0\,.
\end{equation}
There are two other conditions that we have \emph{not} used, but that serve as further consistency checks of the above system of equations:
\begin{enumerate}\setcounter{enumi}{4}
\item We will show below in \eqref{wrong vev gcg} that
$\langle W_{\mathrm{GCG}}\rangle = W - \frac{1}{\pi}\re T \partial_T W_{\mathrm{np}}(2\alpha-\beta-\xi)$.  Hence, if we were to require $\langle W_{\mathrm{GCG}}\rangle = W$, as explained in \S\ref{sssw}, then we would obtain the condition $2\alpha-\beta-\xi=0$, which is fulfilled by \eqref{valuesofcoeffs}.
\item The integrability condition  obtained from \eqref{killing1} is\footnote{In the interest of complete generality, one could have added a smeared correction $2\gamma \langle S\rangle/\mathcal{V}_\perp$ to the right-hand side of \eqref{killing1}.  However, the integrability condition then requires $\alpha+\gamma=1$, whereas the consistency condition from $G_{(1,2)}$ requires $\alpha=1$, and so $\gamma=0$.  We have therefore not included such a term in \eqref{killing1}.}
\begin{equation}
-6i \mu\mathcal{V}_\perp +2\alpha \langle S\rangle =0\,,
\end{equation}
where we used $t=3\mathcal{V}_\perp.$  We use the relation $\mu=-i e^{\phi/2} W/(4\pi \mathcal{V})$ to rewrite the integrability condition as
\begin{equation}
\alpha \partial_T W_{\mathrm{np}} +\kappa_4^2 K_T W=0.
\end{equation}
Hence, we obtain $\alpha=1$, which accords with the above.
\end{enumerate}

In summary, we find that the Killing spinor equations that consistently incorporate the effects of gaugino condensation are:
\begin{align}
d_H \Bigl(e^{(\phi/4-A)\hat{p}}e^{3A-\phi/4}\Phi_2\Bigr) =~&2i \mu e^{(\phi/4-A)\hat{p}}e^{2A-\phi/2}\im\Phi_1+2\langle S\rangle \delta^{(2)}\,,\label{killing1final}\\
d_H\Bigl(e^{(\phi/4-A)\hat{p}}e^{2A-\phi/2}\im\Phi_1\Bigr) =~&0\,,\label{killing2final}\\
d_H\Bigl(e^{(\phi/4-A)\hat{p}}e^{4A}\re\Phi_1\Bigr) =~&3 e^{(\phi/4-A)\hat{p}}e^{3A-\phi/4}\re(\overline{\mu}\Phi_2)+e^{(2A-\phi/2)(3-\hat{p})} e^{4A+\phi}\tilde{F}\label{killing3final}\nonumber\\  &+e^{\phi/2}\re(\langle S\rangle \overline{\Omega})\,\delta^{(0)}\,.
\end{align}
These equations, which differ from those of \cite{Koerber:2007xk,Bena:2019mte}\footnote{The findings in \S6 of \cite{Koerber:2007xk} were arrived at using \eqref{killing3} rather than \eqref{killing3final}, but in many (though not all) respects appear consistent with ours, even though we have used \eqref{killing3final}.  The reason for the near-match is that in \cite{Koerber:2007xk} a nonperturbative superpotential term was \emph{added} to the generalized complex geometry superpotential.  According to our analysis, \eqref{killing3final} should be used, and then no addition is needed, nor indeed would one be consistent.} by the presence of the final term\footnote{This term can also be derived from the results of \cite{Dymarsky:2010mf} (for related approaches, see \cite{Bergshoeff:2001pv,Held:2010az}).  The fluxes we find from \eqref{killing3final}, but not those following from the unmodified \eqref{killing3}, agree with the fluxes obtained in \cite{Dymarsky:2010mf}, after accounting for a difference in normalization.} in \eqref{killing3final}, constitute one of the main results of this Appendix.

\subsubsection{The superpotential}\label{sssw}

In a general type IIB string compactification on a generalized complex geometry, the superpotential is \cite{Grana:2005ny,Benmachiche:2006df,Micu:2007rd,Koerber:2007xk,Lust:2008zd}
\begin{equation}
W_{\mathrm{GCG}}=\pi\int \biggl\langle \Phi_2, \tilde{F}+e^{-4A-\phi}d_H \bigl(e^{4A+(\phi/4-A)\hat{p}}\,\text{Re}\,\Phi_1\bigr)\biggr\rangle\label{eqn:superpotential}\,.
\end{equation}
We will now explain how to evaluate \eqref{eqn:superpotential} in our solution.

In the dynamic SU(2) structure background sourced by gaugino condensation, the one-form and five-form components of $e^{4A+\phi}\tilde{F}+d_H (e^{-\phi+(\phi/4-A)\hat{p}}\,\text{Re}\,\Phi_1)$ and $\Phi_2$, and the $(0,3)$ component of $e^{4A+\phi}\tilde{F}+d_H (e^{-\phi+(\phi/4-A)\hat{p}}\,\text{Re}\,\Phi_1)$, are $\mathcal{O}(\langle\lambda\lambda\rangle).$
However, the three-form component of $\Phi_2$ is $\Omega+\mathcal{O}(\langle\lambda\lambda\rangle^2).$ Hence, collecting the terms
in \eqref{eqn:superpotential} up to order $\mathcal{O}(\langle\lambda\lambda\rangle),$ we obtain the generalized Gukov-Vafa-Witten flux superpotential,
\begin{equation}\label{appthew}
W_{\mathrm{GCG}}=\pi\int \mathfrak{G}\wedge \Omega\,.
\end{equation}
with
\begin{equation}\label{appthepromotion}
\mathfrak{G} :=  G_3+i\mathrm{d}\mathfrak{t}\,.
\end{equation}

For the computations of \S\ref{sec:SUSY} --- in particular, to arrive at \eqref{eqn:ciugcg dim reduction} --- we need to compute $\mathfrak{G}_{0,3}$.
Let us temporarily work with expressions that follow from the general Killing spinor equations \eqref{killing1}-\eqref{killing3} rather than from the particular form \eqref{killing1final}-\eqref{killing3final} that results from imposing \eqref{valuesofcoeffs}.
One can then write \eqref{sol:killing} as
\begin{equation}
\mathfrak{G}_{0,3}=-\frac{e^{-4A}\overline{\Omega}}{\pi\int_M e^{-4A}\Omega\wedge\overline{\Omega}} W +i\frac{e^{-4A-\phi/2}}{4}\langle S\rangle \overline{\Omega}\left((2\alpha-\beta)\delta^{(0)}-\frac{\xi}{\mathcal{V}_\perp}\right).\label{sol:killing2}
\end{equation}
Thus, the vev of the generalized complex geometry superpotential $W_{\mathrm{GCG}}$ on the solution of the ten-dimensional Killing spinor equations is given by
\begin{equation}
\langle W_{\mathrm{GCG}}\rangle = W - \frac{1}{\pi}\re T \partial_T W_{\mathrm{np}}(2\alpha-\beta-\xi)\,,\label{wrong vev gcg}
\end{equation}
so that $\langle W_{\mathrm{GCG}} \rangle = W$ if and only if $2\alpha-\beta-\xi=0$.  On imposing \eqref{valuesofcoeffs} we conclude that
\begin{equation}\label{gcgisfull}
\langle W_{\mathrm{GCG}}\rangle =W\,.
\end{equation}

Using \eqref{appthepromotion}, we can now combine \eqref{sol:killing2} and \eqref{idt} to compute $G_{(0,3)}$:
\begin{equation}
G_{(0,3)}=\underbrace{-\frac{e^{-4A}\overline{\Omega}}{\pi\int_M e^{-4A}\Omega\wedge\overline{\Omega}}W}_{~~~~~~=: G_W} \underbrace{-\frac{i}{64\pi^2}e^{-4A-\phi/2}\langle\lambda\lambda\rangle \overline{\Omega}\left(\delta^{(0)}-\frac{1}{\mathcal{V}_\perp}\right)}_{~~~~~~=: G_{\lambda\lambda}}\,.\label{eqn:flux2}
\end{equation}

Our result accords with \cite{Moritz:2017xto}, where it was shown that in the presence of gaugino condensation (and upon converting to our normalizations), one has
\begin{equation}
G_{(0,3)}=-\frac{i}{64\pi^2}e^{-4A-\phi/2}\langle\lambda\lambda\rangle \overline{\Omega}\delta^{(0)}+G_0\,,\label{eqn:flux}
\end{equation}
for some $G_0$ with $d G_0=0.$  Thus we find agreement between \cite{Moritz:2017xto} and the singular term in \eqref{eqn:flux2}, and moreover we learn that $G_0$ is given by the nonsingular terms in \eqref{eqn:flux2}.

\subsection{Dimensional reduction and translation to four-dimensional terms}\label{appto4d}

We will now use the results of \S\ref{wgravapp} to compute the four-dimensional potential terms that result from dimensional reduction of the gaugino-flux coupling \eqref{eqn:the flux-gaugino coupling} and the four-gaugino term \eqref{finalfourg}, upon assigning the gaugino bilinear vev \eqref{eqn:gaugino vev}.

In our specific setup, $\mathfrak{t}$ is sourced only by gaugino condensation on $D,$ and is given by \eqref{eqn:mathfrakt}.
Writing $\re\Omega_2=\frac{1}{2}\left(\Omega_{12}dz^1\wedge dz^2+\overline{\Omega}_{\bar{1}\bar{2}}d\bar{z}^{\bar{1}}\wedge d\bar{z}^{\bar{2}}\right),$ we have
\begin{equation}\label{d2mfis}
d_2\mathfrak{t}=-\frac{1}{2}\partial_a (e^{\phi/2-2A}\varphi\overline{\Omega}_{\bar{1}\bar{2}})d\bar{z}^{\bar{1}}\wedge d\bar{z}^{\bar{2}}\wedge dz^a-\frac{1}{2}\partial_{\bar{a}} (e^{\phi/2-2A}\varphi\Omega_{12})dz^{1}\wedge dz^{2}\wedge d\bar{z}^{\bar{a}}.
\end{equation}
It follows from the index structure of \eqref{d2mfis} that $d_2\mathfrak{t}\cdot\Omega=0$.
Thus we arrive at
\begin{equation}\label{noidt}
\int_{X\times D} \sqrt{-g} \,d_2\mathfrak{t}\cdot \Omega =0\,.
\end{equation}
Assigning the gaugino bilinear vev \eqref{eqn:gaugino vev} and using \eqref{noidt} and \eqref{eqn:flux2}, the coupling \eqref{eqn:the flux-gaugino coupling} dimensionally reduces to
\begin{align}
S_{\mathfrak{G}\lambda\lambda}=&-\int_{X\times M}\sqrt{-g}e^{\phi/2-6u}\frac{ie^{-4A+4u}\Omega\cdot\overline{\Omega}}{\pi \int_M e^{-4A}\Omega\wedge\overline{\Omega}}\frac{\langle\bar{\lambda}\bar{\lambda}\rangle}{32\pi}W\delta^{(0)}+c.c.+S_{\lambda\lambda}^{\mathrm{sing}}, \label{eq57} \\
=&\int_X\sqrt{-g_4}e^{\phi/2-6u+\kappa_4^2K/2}\frac{\text{Re}(T)}{2\pi\mathcal{V}}\diff_{\overbar{T}}\overline{W}W+c.c.+S_{\lambda\lambda}^{\mathrm{sing}},\\
=&-\kappa_4^2 \int_X \sqrt{-g_4}e^{\kappa_4^2K}K^{T\overbar{T}}\diff_{\overbar{T}}\overline{W} K_T W+c.c.+S_{\lambda\lambda}^{\mathrm{sing}}\,, \label{llsing}
\end{align}
where the singular term
\begin{equation}\label{ssingis}
S_{\lambda\lambda}^{\mathrm{sing}}=\frac{i}{32\pi}\int \sqrt{-g}e^{\phi/2-2u}G_{\lambda\lambda}\cdot \Omega\bar{\lambda}\bar{\lambda}\delta^{(0)}+c.c.\,,
\end{equation} with $G_{\lambda\lambda}$ given in \eqref{eqn:flux2}, is analyzed in Appendix \ref{app:d3}.
We used the identity $\kappa_4^2 K^{T\overbar{T}}K_T=-\text{Re}(T)/(2\pi\mathcal{V})$, which follows from \eqref{kpotis} and \eqref{thevis}.

Similarly, assigning the gaugino bilinear vev \eqref{eqn:gaugino vev}, the integral of the four-gaugino term \eqref{finalfourg} dimensionally reduces to
\begin{align}\label{eqn:fourgauginofinalfull}
S_{\lambda\lambda\lambda\lambda}=&-\int_X \int_M \sqrt{-g} e^{\kappa_4^2 K+4u} \frac{e^{-4A+4u}\Omega\cdot \overline{\Omega}}{24\pi \mathcal{V}_\perp}  \diff_T W_{\mathrm{np}} \diff_{\bar{T}}\overline{W}_{\mathrm{np}}\delta^{(0)}\\
=&-\int_X \sqrt{-g_4}e^{\kappa_4^2K} \frac{\text{Re}(T)^2}{3\pi\mathcal{V}} \diff_T W\diff_{\bar{T}}\overline{W}\\
=&-\int_X \sqrt{-g_4}e^{\kappa_4^2K}K^{T\bar{T}}\diff_T W\diff_{\bar{T}}\overline{W}.
\end{align}
We used the identity $K^{T\overbar{T}}=\text{Re}(T)^2/(3\pi \mathcal{V})$.

The \emph{modified} Killing spinor equations \eqref{killing1final}-\eqref{killing3final} were crucial in the above: if instead of \eqref{gcgisfull} one had
$\langle W_{\mathrm{GCG}} \rangle \overset{!?}{=} W_{\mathrm{flux}}$ then in \eqref{llsing} the factor $K_T W$ would instead read $K_T W_{\mathrm{flux}}$, and the scalar potential obtained from ten dimensions would disagree with that obtained in four-dimensional supergravity.  However, we reiterate that the form \eqref{killing1final}-\eqref{killing3final} of the Killing spinor equations was not derived by requiring that they should lead to \eqref{gcgisfull}; instead, the logically independent consistency conditions of \S\ref{sssappsumm} were imposed to derive \eqref{killing1final}-\eqref{killing3final}, and \eqref{gcgisfull} was then a consequence.\footnote{Although \eqref{gcgisfull} is essential to our derivation of the correct finite four-dimensional potential \eqref{10dans} from a ten-dimensional configuration, the cancellation of divergences exhibited in Appendix \ref{app:d3} does \emph{not} rely on \eqref{gcgisfull}.}

\subsection{Normalization of the K\"ahler potential}\label{kapp}

We temporarily normalize the flux superpotential as
\begin{equation}
W_{\mathrm{flux}}= a\int_M G\wedge\Omega\,,
\end{equation}
and the K\"ahler potential as
\begin{equation}
\kappa_4^2 K=-3\log\bigl(T+\overbar{T}\bigr)-\log\left(i \int_M e^{-4A} \Omega\wedge\overline{\Omega}\right)-\log\bigl(-i(\tau-\overbar{\tau})\bigr)-\log b\,.
\end{equation}
Given a complex structure, we normalize
\begin{equation}
i \int_M e^{-4A}\Omega\wedge\overline{\Omega}=c\,.
\end{equation}
We now fix $a,$ $b,$ and $c$ by dimensional reduction of the ten-dimensional supergravity action.

The first constraint is given by matching the F-term potential for the complex structure moduli and axiodilaton.  Matching the gravitino mass does not provide an additional constraint.
The potential
\begin{align}
V_\tau= &\frac{1}{2\kappa_{10}^2}\int_M\sqrt{g_6} e^{4A-12u+\phi}|G_{3,0}|^2\\
=&\frac{1}{2\kappa_{10}^2}\int_Me^{4A-12u+\phi} \left(\frac{ \int_M G\wedge \overline{\Omega}}{\int_M e^{-4A} \Omega\wedge\overline{\Omega}}e^{-4A}\Omega \right) \wedge \star_6 \left(-\frac{\int_M \overbar{G}\wedge\Omega}{\int_M e^{-4A}\Omega\wedge\overline{\Omega}}e^{-4A}\overline{\Omega} \right)\\
=&\frac{1}{2\kappa_{10}^2}e^{-12u+\phi} \frac{\int_M G\wedge\overline{\Omega} \int_M \overbar{G}\wedge\Omega}{i \int_M e^{-4A}\Omega\wedge\overline{\Omega}}
\end{align}
must match
\begin{align}
V_\tau= e^{\kappa_4^2 K} K^{\tau\overbar{\tau}} D_\tau WD_{\overbar{\tau}}\overline{W}
=\kappa_4^2 e^{\kappa_4^2 K}a^2 \int_M G\wedge\overline{\Omega}\int_M \overline{G}\wedge\Omega\,,
\end{align}
which requires
\begin{equation}
\frac{a^2}{b}=2^7\pi^2 \mathcal{V}^3\,.
\end{equation}

Another constraint is given by matching the F-term potential for D3-brane moduli.
Matching the F-term potential for the K\"ahler modulus does not provide an additional constraint.  From \eqref{eqn:potential from flux} with the undetermined coefficient $c$ we have
\begin{equation}
\Phi_{-}=c\frac{e^{\kappa_4^2 K}}{8\mu_3 \mathcal{V}} K^{a\bar{b}}D_aWD_{\bar{b}}\overline{W}\,.
\end{equation}
Hence we fix
\begin{equation}
i\int_M e^{-4A}\Omega\wedge\overline{\Omega}=8\mathcal{V}\,.
\end{equation}

There remains the freedom to choose $a$ and $b$, corresponding to K\"ahler invariance.
All such choices are physically equivalent; for the sake of simplicity we normalize the superpotential as
\begin{equation}
\pi\int_M G\wedge\Omega\,,
\end{equation}
and the K\"ahler potential as
\begin{equation}
\kappa_4^2K=-3\log\bigl(T+\overbar{T}\bigr) -\log\left(i\int_Me^{-4A}\Omega\wedge\overline{\Omega}\right)-\log\bigl(-i(\tau-\overbar{\tau})\bigr)+\log\bigl(2^7\mathcal{V}^3\bigr)\,.
\end{equation}

\section{Spectroscopy of Interactions} \label{app:spec}

In this appendix we show that the interactions of anti-D3-branes with a gaugino condensate that are mediated by Kaluza-Klein excitations of a Klebanov-Strassler throat can be safely neglected, in the sense defined in \S\ref{sec:dS}.

\subsection{Kaluza-Klein modes on $T^{1,1}$}

We will use the conventions of \cite{Baumann:2010sx} for denoting fields on the conifold and operators in the Klebanov-Witten theory.
We use labels $L \equiv (j_1,j_2,R)$ and $M \equiv (m_1,m_2)$ for the quantum numbers under the $SU(2)\times SU(2)\times U(1)_R$ isometries of $T^{1,1}$, and write a solution to the Laplace equation on the conifold, $\nabla^2 f = 0$, as
\begin{equation}\label{harmonicfn}
f(r,\Psi) = \sum_{L,M} f_{LM}\Bigl(\frac{r}{r_{\mathrm{UV}}}\Bigr)^{\Delta_s(L)}Y_{LM}(\Psi)\,,
\end{equation} with the eigenvalues\footnote{The eigenvalues $\Delta_s(L)$ were denoted by $\Delta(L)$ in \cite{Baumann:2008kq}, by $\Delta_f(L)$ in \cite{Baumann:2010sx}, and by $\Delta(I_s)-4$ in \cite{Gandhi:2011id}.}
\begin{equation}\label{harmonicfneig}
\Delta_s(L) = -2 + \sqrt{6\bigl[j_1(j_1+1)+j_2(j_2+1)-R^2/8\bigr]+4}\,.
\end{equation} The singlet $j_1=j_2=R=0$ has $\Delta_s=0$, and the next-lowest eigenvalue, for $j_1=j_2=1/2, R=1$, is $\Delta_s=3/2$.

\subsubsection{Perturbations sourced by D3-branes and anti-D3-branes}

We now consider in turn the perturbations sourced by D3-branes or anti-D3-branes in the infrared or ultraviolet regions of a Klebanov-Strassler throat.
Recall that the  Dirac-Born-Infeld + Chern-Simons action of a probe D3-brane is $S_{D3}=\mu_3 \Phi_-$,
and a D3-brane is a localized source for the scalar $\Phi_+$, whereas
the Dirac-Born-Infeld + Chern-Simons action of a probe anti-D3-brane is
$S_{\overbar{D3}}=\mu_3 \Phi_+$, and an anti-D3-brane is a localized source for the scalar $\Phi_-$.
As explained in \cite{Baumann:2008kq}, see also \cite{Gandhi:2011id}, it is useful to define the fields $\varphi_+:=r^{4}\Phi_+^{-1}$ and $\varphi_-:=r^{-4}\Phi_-$,
which have canonical kinetic terms and so have solutions of the usual form
\begin{equation}\label{nowcanon}
\varphi_\pm = \alpha\, r^{-\Delta_\pm} + \beta\, r^{\Delta_\pm-4}\,.
\end{equation} with $\alpha$, $\beta$ independent of $r$.

\begin{itemize}

\item{}Anti-D3-brane in the infrared:

The leading perturbation of $\Phi_-$ is a normalizable profile,
\begin{equation}\label{d3bargives1}
\delta\Bigl(r^{-4}\Phi_-\Bigr) \sim r^{-8-\Delta_s(L)}\,.
\end{equation}
The leading (singlet) mode scales as $r^{-8}$,
and corresponds in the dual field theory to an expectation value for the dimension-eight operator \cite{Kachru:2007xp,Baumann:2008kq,Dymarsky:2011pm}
\begin{equation}\label{o8is}
\mathcal{O}_8 = \int d^2\theta d^2\bar{\theta}\,\mathrm{Tr}\bigl[W_+^2 \overline{W}_+^2\bigr]\,.
\end{equation}
Higher multipoles in the linear solution result from operators such as (but not limited to, cf.~\cite{Baumann:2008kq,Baumann:2010sx})
\begin{equation}\label{o83k2is}
\mathcal{O}_{8+3k/2} = \int d^2\theta d^2\bar{\theta}\,\mathrm{Tr}\bigl[W_+^2 \overline{W}_+^2\,\bigr (AB)^k]\,,
\end{equation}
for $k \in \mathbb{Z}_+$.
The first non-singlet mode is $\mathcal{O}_{19/2}$, and scales as $r^{-19/2}$.
See \cite{Baumann:2008kq,Baumann:2010sx,Gandhi:2011id} for extensive analysis of this system.

\item{}D3-brane in the infrared:

The leading perturbation of $\Phi_+$ is a normalizable profile,
\begin{equation}\label{d3gives}
\delta\Bigl(r^{4}\Phi_+^{-1}\Bigr) \sim r^{-\Delta_s(L)}\,.
\end{equation}
The singlet is a constant, while higher multipoles correspond to expectation values for operators such as (but not limited to, cf.~\cite{Baumann:2008kq})
\begin{equation}\label{o3k2is}
\mathcal{O}_{3k/2} = \mathrm{Tr}\bigl[(AB)^k\bigr]\bigr\vert_b\,,
\end{equation}
for $k \in \mathbb{Z}_+$, with $\vert_b$ denoting the bottom ($\theta=\bar{\theta}=0$) component of a supermultiplet, as in \cite{Baumann:2010sx}.
The leading non-singlet mode scales as $r^{-3/2}$ \cite{Baumann:2006th,Baumann:2008kq,Baumann:2010sx,Gandhi:2011id},
and is dual to an expectation value for
\begin{equation}\label{o32is}
\mathcal{O}_{3/2} = \mathrm{Tr}\bigl[AB\bigr]\bigr\vert_{b}\,.
\end{equation}
Higher multipoles can be found in \cite{Baumann:2008kq,Baumann:2010sx,Gandhi:2011id}.

\item{}D3-brane in the ultraviolet:

The leading perturbation of $\Phi_+$ is a non-normalizable profile \cite{Baumann:2008kq}
\begin{equation}\label{d3uvgives}
\delta\Bigl(r^{4}\Phi_+^{-1}\Bigr) \sim r^{\Delta_s(L)+4}\,.
\end{equation}
The singlet mode scales as $r^4$,
and is dual to a source for the operator $\mathcal{O}_{8}$ in \eqref{o8is}
whose expectation value arose in the anti-D3-brane solution \eqref{d3bargives1}.
Higher multipoles are dual to sources for operators such as $\mathcal{O}_{8+3k/2}$ in \eqref{o83k2is}.
The leading non-singlet mode scales as $r^{11/2}$, and is dual to $\mathcal{O}_{19/2}$
\cite{Baumann:2008kq,Baumann:2010sx,Gandhi:2011id}.

\end{itemize}

\subsection{Effect of anti-D3-branes on gaugino condensate}\label{d3baroncond}

We would like to examine the long-distance solution sourced by $p$ anti-D3-branes smeared\footnote{At different stages of the evolution of a collection of anti-D3-branes interacting with flux, as described in \cite{Kachru:2002gs}, the anti-D3-branes may be localized at a point on the $S^3$ at the tip, or puffed up into a nontrivial configuration, and in such a case the supergravity equations of motion become difficult partial differential equations.  Fortunately (cf.~\cite{Dymarsky:2011pm}), in any of these cases the leading long-distance solution linearized around $AdS_5 \times T^{1,1}$ can be obtained from the $SU(2)\times SU(2)$ invariant part of the linearized solution, i.e.~from the linearized solution obtained from considering anti-D3-branes smeared around the $S^3$. This latter problem requires solving only ordinary differential equations.} around the tip of a Klebanov-Strassler throat.
To start out, we will linearize in the strength of the anti-D3-brane backreaction, and then discuss
nonlinear effects.

\subsubsection{Coulomb interaction with a D3-brane}

The $SU(2)\times SU(2)$ invariant part of the linearized long-distance solution sourced by $p$ anti-D3-branes at the tip of a noncompact Klebanov-Strassler throat has been studied in \cite{DeWolfe:2008zy,Bena:2009xk,Bena:2010ze,Bena:2011hz,Dymarsky:2011pm,Bena:2011wh,Dymarsky:2013tna}.
The leading perturbation of $\Phi_-$ corresponds to the normalizable profile \eqref{d3bargives1}, up to logarithmic corrections.

A strong consistency check of this solution comes from considering a D3-brane in the ultraviolet region of the throat.
The potential for motion of such a D3-brane can be computed either by treating the D3-brane as a probe in the solution \eqref{d3bargives1} sourced by the anti-D3-branes,
or by treating the anti-D3-branes as probes in the solution sourced by the backreaction of a D3-brane in a Klebanov-Strassler throat.
The former approach amounts to evaluating the action
of a probe D3-brane in the solution of
\cite{Bena:2009xk,Bena:2010ze,Bena:2011hz,Dymarsky:2011pm,Bena:2011wh}.

The latter approach, which was used to compute the D3-brane Coulomb potential in \cite{Kachru:2003sx}, is even simpler, because the D3-brane and the Klebanov-Strassler background preserve the same supersymmetry, and so the perturbation due to the D3-brane enjoys harmonic superposition.
One finds \cite{Baumann:2008kq} that
the leading perturbation of $\Phi_+$
sourced by D3-brane in the ultraviolet is the non-normalizable profile \eqref{d3uvgives}.

The Coulomb potential between an anti-D3-brane in the infrared and a D3-brane in the ultraviolet can be computed either from \eqref{d3bargives1} \cite{Baumann:2008kq,Bena:2010ze} or from \eqref{d3uvgives} \cite{Kachru:2003sx}, with exact agreement.

We can understand this match in the language of the dual field theory (see \S3.3 of \cite{Baumann:2008kq}).
A D3-brane in the ultraviolet creates a potential by sourcing a non-normalizable\footnote{In the sense of footnote 8 of \cite{Baumann:2008kq}.} profile $\delta\Phi_+$, corresponding to a \emph{source} (in the field theory Lagrangian) for operators such as $\mathcal{O}_{8}$.
An anti-D3-brane in the infrared creates a potential by sourcing a normalizable profile $\delta\Phi_-$, corresponding to an \emph{expectation value} for operators such as $\mathcal{O}_{8}$.
Either way, the mediation occurs by a high-dimension operator, and leads to a very feeble interaction at long distances.

The above arguments give several conceptually different --- but precisely compatible --- perspectives on a single fact, which is that the Coulomb interaction of a D3-brane with an anti-D3-brane in a warped region is suppressed by eight powers of the warp factor, and so is extremely weak \cite{Kachru:2003sx}.

\subsubsection{D3-brane perturbation to gauge coupling}\label{d3pertgc}

Thus far, as a first step, we have used a D3-brane in the ultraviolet as a probe of the solution generated by anti-D3-branes in the infrared.
Our actual interest is in the effect of anti-D3-branes in the infrared on D7-branes in the ultraviolet.

Now, as a further warm-up, we recall the effect of D3-branes (not yet anti-D3-branes) in the infrared on gaugino condensation on D7-branes in the ultraviolet.\footnote{Corrections to gaugino condensation on D7-branes due to interactions with distant branes have been extensively studied in the context of D3-brane inflation,
both from the open string worldsheet \cite{Berg:2004ek,Berg:2004sj} and in supergravity \cite{Baumann:2006th}: see \cite{Baumann:2014nda} for a review.}
The effect of the perturbation \eqref{d3gives} on a gaugino condensate was computed in \cite{Baumann:2006th}.
Upon summing over all the chiral and non-chiral operators of the Klebanov-Witten theory \cite{Klebanov:1998hh}, and applying highly nontrivial identities to collapse the sum,
the result for $\delta T$ took the form of a logarithm of the embedding function of the D7-branes, expressed in local coordinates \cite{Baumann:2006th}.
The perturbation \eqref{d3gives} is thus the effect responsible for the dependence of the gaugino condensate on the D3-brane position  \cite{Berg:2004ek,Baumann:2006th}, which is of central importance in D3-brane inflation \cite{Kachru:2003sx}.

This result was exactly reproduced by an entirely different computation in \cite{Baumann:2010sx}, as reviewed in Appendix \ref{app:d3} below: the $G_-$ flux sourced by the gaugino-flux couplings on the D7-branes leads to a solution for $\Phi_-$, and a D3-brane probing this solution experiences the potential implied by the perturbation $\delta T$ computed in \cite{Baumann:2006th}.

For completeness, we now explain an asymmetry between the effects of D3-branes and of anti-D3-branes.
As will be explained in \S\ref{d3barcoupl} below, one finds
from \eqref{d3bargives1} that an anti-D3-brane in the infrared has only extremely small effects on D3-branes or D7-branes in the ultraviolet (except through couplings via the zero-mode $e^u$).
In contrast, a D3-brane in the infrared \emph{does} have a detectable effect at long distances.
Adding a D3-brane increases the total D3-brane charge of the throat by one unit, $N \to N+1$,
and this change is reflected in the solution by a non-normalizable correction relative to the throat with $N$ units of flux and no D3-brane.

Simply adding an anti-D3-brane would likewise change the net tadpole and the flux, and so have a detectable effect at long distances.
However, this is not the relevant comparison for our purposes.
The anti-D3-brane configuration of \cite{Kachru:2002gs} is a metastable state in a throat with less flux and some wandering D3-branes, but the \emph{same} total tadpole.
The anti-D3-branes thus source small normalizable corrections to the solution that is dual to the supersymmetric ground state.

\subsubsection{Anti-D3-brane perturbation to gauge coupling}\label{d3barcoupl}

To compute the effect on the gaugino condensate of the perturbation \eqref{d3bargives1} due to anti-D3-branes in the infrared, we follow the same logic used in \cite{Baumann:2006th} and reviewed in \S\ref{d3pertgc}.
We evaluate the D7-brane gauge coupling function \eqref{deft},
\begin{equation}\label{deftrecap1}
T=e^{4u} \int_D \sqrt{g_6} e^{-4A} +i\int_D C_4\,,
\end{equation} in the perturbed solution, and use \eqref{wnpis}.
Examining \eqref{deftrecap1}, we see that it suffices to know the breathing mode $e^{u}$, as well as the leading perturbations to $\Phi_{\pm}$ and to the metric $g_{ab}$ at the location of the D7-brane.  Because $e^{u}$ is a six-dimensional zero-mode, we will treat it separately: at this stage we seek to check that any influences of the anti-D3-branes on the condensate, \emph{except} via the breathing mode, can be neglected.

Because $\Phi_-=0$ in the Klebanov-Strassler background, we write (see Appendix D of \cite{McAllister:2016vzi})
\begin{equation}\label{deftrecap2}
\delta\,\mathrm{Re}\,T \approx  e^{4u} \int_D \sqrt{g_{(0)}}\Biggl(-2\bigl(\Phi_{+}^{(0)}\bigr)^{-2}\bigl(\delta\Phi_+ + \delta\Phi_-\bigr) + \bigl(\Phi_{+}^{(0)}\bigr)^{-1} g^{ab}_{(0)}\delta g_{ab}\Biggr)\,,
\end{equation}
where for a field $\phi$, the background profile in the Klebanov-Strassler solution is denoted $\phi^{(0)}$.

Our consideration above of a D3-brane probe in the ultraviolet showed that $\delta\Phi_-$ is mediated by $\mathcal{O}_8$ (with subleading corrections from operators of even higher dimension) and is negligible at the D7-brane location.
Perturbations $\delta\Phi_+$ (or more usefully, $\delta\varphi_+$) are mediated by operators such as $\mathcal{O}_{3/2}$, and can be sizable \emph{if strongly sourced}, e.g.~by the presence of a D3-brane.  However, in \cite{Gandhi:2011id} it was shown that the leading profile $\delta\varphi_+$ that arises in the full nonlinear solution due to an anti-D3-brane scales as $\delta\varphi_+ \sim r^{-8}$, just like the profile $\delta\varphi_-$ in \eqref{d3bargives1} that is directly sourced by the anti-D3-brane: see \S5 of \cite{Gandhi:2011id}.  Likewise, in Appendix D of \cite{McAllister:2016vzi} it was shown that the leading non-singlet metric perturbation scales as $r^{-19/2}$ (see \cite{Gandhi:2011id,McAllister:2016vzi} for definitions of the associated tensor harmonics on $T^{1,1}$).

In summary, in the linearized background \eqref{d3bargives1} sourced by anti-D3-branes in the infrared, the leading corrections to $\mathrm{Re}\,T$
are mediated by operators of dimension $\Delta \ge 8$, resulting in extremely small corrections to the D7-brane gaugino condensate when the hierarchy of scales in the Klebanov-Strassler throat is large.
Thus, the only influence of the anti-D3-branes on the gaugino condensate that is non-negligible for our purposes occurs via the breathing mode $e^u$, and was already included in the four-dimensional analysis of \cite{KKLT}.  We have therefore established \eqref{thetotTapproxyes}.

\subsection{Effect of gaugino condensate on anti-D3-branes}

For the avoidance of doubt, we now reverse the roles of source and probe relative to \S\ref{d3baroncond}, and examine the influence of gaugino condensation in the ultraviolet on anti-D3-branes in the infrared.  As in \S\ref{d3baroncond}, we treat the breathing mode separately.

\subsubsection{Leading effect of flux}\label{sss:leof}

The anti-D3-brane probe action is $S_{\overbar{D3}}=\mu_3 \Phi_+$, so we seek the leading perturbations of $\Phi_+$ in the infrared.
Gaugino condensation on D7-branes directly sources flux perturbations $\delta G_-$ and $\delta G_+$ via the gaugino-flux coupling \eqref{ciunogcg}, as shown in \cite{Baumann:2010sx} and reviewed in \S\ref{sec:SUSY}.
Expanding in Kaluza-Klein modes on $T^{1,1}$, the lowest mode of $\delta G_+$ is dual to the operator
\begin{equation}\label{o52is}
\mathcal{O}_{5/2} = \int d^2 \theta\, \mathrm{Tr}\bigl[AB\bigr]\,,
\end{equation} of dimension $\Delta=5/2$ \cite{Baumann:2010sx}.
The coefficient $c_{5/2}$ of this mode in the ultraviolet is at most of order $\langle\lambda\lambda\rangle$, because it is incompatible with the no-scale symmetry of the Klebanov-Strassler background, and so is present only once it is sourced by the gaugino condensate \cite{Baumann:2010sx,Gandhi:2011id}.  We stress, however, that $c_{5/2}$
might well be parametrically smaller than $\langle\lambda\lambda\rangle$: the operator $\mathcal{O}_{5/2}$ is easily forbidden by (approximate) symmetries, corresponding in the bulk to symmetries of the D7-brane configuration.\footnote{See e.g.~\cite{Kachru:2009kg} for related work.}  Our estimates of the anti-D3-brane potential will therefore be upper bounds.

The equation of motion for the scalar $\Phi_+$ is
\begin{equation}
\nabla^2\Phi_+=\frac{e^{8A}}{\im\tau}|G_+|^2 + \ldots \label{eqn:phi plus from flux}
\end{equation} where the omitted terms (cf.~\S\ref{sec:EOM}) can be neglected for the present purpose.
In the Klebanov-Strassler background, the three-form flux has a nonvanishing profile $G_+^{(0)}$ \cite{Klebanov:2000hb}.
With one insertion of the background flux and one insertion of the perturbation $\delta G_+$, we have
\begin{equation}
\nabla^2\Phi_+=\frac{e^{8A}}{\im\tau} \Bigl(G_+^{(0)}\cdot \delta G_++c.c.\Bigr)\,, \label{eqn:phi plus from flux pert}
\end{equation} from which one finds
\begin{equation}
\delta\Phi_+ \sim e^{\frac{5}{2}A_{\mathrm{tip}}} \times \langle\lambda\lambda\rangle\,,
\end{equation} with $e^{A_{\mathrm{tip}}}$ the warp factor at the tip.
Since
\begin{equation}\label{langlewarp}
\langle\lambda\lambda\rangle \sim \mathcal{O}(e^{2A_{\mathrm{tip}}})\,,
\end{equation}
we conclude that
\begin{equation} \label{delvfromgplus}
\delta V_{\overbar{D3}} \lesssim \mu_3 e^{\frac{9}{2}A_{\mathrm{tip}}}\,,
\end{equation} which is smaller, by a power $e^{\frac{1}{2}A_{\mathrm{tip}}}$, than the anti-D3-brane potential \eqref{eqn:stress energy D3} in the Klebanov-Strassler background.
Thus, the influence of the gaugino condensate on the anti-D3-brane, via the linearized perturbation $\delta G_+$, is a parametrically small correction.

\subsubsection{Spurion analysis}

Thus far we have considered only the linearized perturbation $\delta G_+$ dual to $\mathcal{O}_{5/2}$, leading to the small correction \eqref{delvfromgplus} to the anti-D3-brane potential.
If the D7-brane configuration enjoys no additional symmetries that enforce $c_{5/2} \ll \langle\lambda\lambda\rangle$, then \eqref{delvfromgplus} is indeed the parametrically dominant correction to the anti-D3-brane potential from gaugino condensation \cite{gandhisjors}.
However, establishing this requires extending the treatment of \S\ref{sss:leof} to incorporate more general perturbations, such as perturbations of the metric, and also requires working
at nonlinear order in these perturbations.  A complete analysis of this system is carried out in \cite{gandhisjors}; here we review the strategy and summarize the main findings.

To find the general form of the infrared solution created by a partially-known ultraviolet source, one can perform a \emph{spurion analysis},
in which the parametric size of the ultraviolet coefficient $c_{\Delta}$ of a given mode $\delta\phi_{\Delta}$ dual to a source for an operator $\mathcal{O}_{\Delta}$
is determined by the symmetries preserved by $\mathcal{O}_{\Delta}$.

Specifically, perturbations allowed in a no-scale compactification of the Klebanov-Strassler throat, as in \cite{Giddings:2001yu}, have $c_{\Delta} \sim \mathcal{O}(1)$.
Perturbations that are allowed only after (a single) insertion of the gaugino condensate expectation value $\langle\lambda\lambda\rangle$ have $c_{\Delta} \sim \mathcal{O}(\langle\lambda\lambda\rangle)$,
while perturbations that are allowed only after inserting $|\langle\lambda\lambda\rangle|^2$ have $c_{\Delta} \sim \mathcal{O}(\langle\lambda\lambda\rangle^2)$.

To determine the spurion assignment for a given operator, we examine couplings of the field theory dual to the throat to the D7-brane field theory.
Consider, for example,
\begin{equation}\label{spur1}
\int d^2 \theta\, \mathrm{Tr}\bigl[AB\bigr]\,\mathrm{Tr}\bigl[W_{\alpha}W^{\alpha}\bigr]_{\mathrm{D7}}\,,
\end{equation} where\footnote{The D7-brane gauge field strength superfield $W^{\alpha}\vert_{\mathrm{D7}}$ should not be confused with $W_+$ appearing in \eqref{o8is}, which is the gauge field strength superfield of the D3-brane fields of the Klebanov-Witten theory.}
\begin{equation}
\Bigl\langle \mathrm{Tr}\bigl[W_{\alpha}W^{\alpha}\bigr]_{\mathrm{D7}} \Bigr\rangle\Bigr\vert_{b} = \frac{1}{2}\langle\lambda\lambda\rangle\,.
\end{equation}
From \eqref{spur1} we find the coupling
\begin{equation}\label{spur1gives}
\delta W = \frac{1}{2}\langle\lambda\lambda\rangle\,\int d^2 \theta\, \mathrm{Tr}\bigl[AB\bigr]\,,
\end{equation} which can be interpreted as a perturbation to the superpotential of the Klebanov-Witten theory, with the exponentially small spurion coefficient $\langle\lambda\lambda\rangle$.

Evidently, to carry out such a spurion analysis one needs to know which perturbations of the supergravity fields are allowed in the background, versus requiring either one or two factors of $\langle\lambda\lambda\rangle$ as spurion coefficients.
This information can be read off from an assignment of the operators of the dual field theory to supermultiplets, as in \cite{Ceresole:1999zs,Ceresole:1999ht}.
A systematic treatment along these lines appears in \cite{Baumann:2010sx,Gandhi:2011id,gandhisjors}.

Examining \eqref{eqn:phi plus from flux}, one sees that the leading linearized perturbations to the anti-D3-brane potential are modes of the flux $G_+$, the axiodilaton $\tau$, and the metric $g$.
At this stage we need to know, from Kaluza-Klein spectroscopy and from spurion analysis, the dimensions $\Delta_{\mathrm{min}}$ of the lowest-dimension non-singlet modes of $G_+$, $\tau$, and $g$, as well as their spurion coefficients $c_{\Delta}$.
For the flux, one finds \cite{gandhisjors}
\begin{equation}\label{leadG}
\Delta_{\mathrm{min}}(G_+) = 5/2\qquad \mathrm{with}\qquad c_{5/2} \sim \langle\lambda\lambda\rangle\,,
\end{equation} corresponding to $\mathcal{O}_{5/2}$ in \eqref{o52is}, as explained above.
Another mode of flux gives a slightly smaller contribution:
\begin{equation}\label{subleadG}
\Delta(G_+) = 3 \qquad \mathrm{with}\qquad c_{3} \sim \langle\lambda\lambda\rangle\,,
\end{equation} corresponding to the operator $\mathcal{O}_{3,+}=\mathrm{Tr}\bigl[W_+^2\bigr]\bigr\vert_b$.
For the dilaton, one finds \cite{gandhisjors}
\begin{equation}\label{leadtau}
\Delta_{\mathrm{min}}(\tau) = 11/2\qquad \mathrm{with}\qquad c_{11/2} \sim \mathcal{O}(1)\,,
\end{equation} corresponding to
\begin{equation}\label{o112is}
\mathcal{O}_{11/2} = \int d^2\theta\,\mathrm{Tr}\bigl[W_+^2\,(AB)\bigr]\,,
\end{equation} which is allowed in the background of \cite{Giddings:2001yu}.  (There is also a $\Delta=4$ mode of $\tau$, but we can absorb this into the background value of the dilaton.)
For the metric, one finds the leading contribution \cite{gandhisjors,Moritz:2017xto}
\begin{equation}\label{leadg}
\Delta_{\mathrm{min}}(g) = 3 \qquad \mathrm{with}\qquad c_{3} \sim \langle\lambda\lambda\rangle\,,
\end{equation} corresponding to
\begin{equation}\label{o3is}
\mathcal{O}_{3,-} = \mathrm{Tr}\bigl[W_-^2\bigr]\bigr\vert_b.
\end{equation}
The first subleading correction from a metric mode has
\begin{equation}\label{subleadg}
\Delta(g) = \sqrt{28} \approx 5.29 \qquad \mathrm{with}\qquad c_{\sqrt{28}} \sim \mathcal{O}(1)\,,
\end{equation} corresponding to
\begin{equation}\label{o529is}
\mathcal{O}_{\sqrt{28}} = \int d^2\theta\,d^2\bar{\theta}\,\mathrm{Tr}\bigl[f(A,B,\bar{A},\bar{B})\bigr]\,,
\end{equation} where $f$ is a harmonic, but not holomorphic, function of the chiral superfields $A$ and $B$.
The perturbation dual to $\mathcal{O}_{\sqrt{28}}$ is allowed in the background of \cite{Giddings:2001yu}.

Using \eqref{langlewarp}, we find from the linearized perturbations \eqref{leadG},\eqref{subleadG},\eqref{leadtau},\eqref{leadg},\eqref{subleadg}
that the anti-D3-brane potential receives corrections of the parametric form
\begin{equation} \label{delvfromall}
\delta V_{\overbar{D3}} \lesssim \mu_3 e^{4 A_{\mathrm{tip}}}\Biggl( e^{\frac{1}{2}A_{\mathrm{tip}}} + e^{A_{\mathrm{tip}}} + e^{(\sqrt{28}-4)A_{\mathrm{tip}}} + e^{\frac{3}{2}A_{\mathrm{tip}}}+\ldots\Biggr)\,.
\end{equation}
For completeness, we remark that upon applying the methods of \cite{Gandhi:2011id} to study the nonlinear solution, one finds \cite{gandhisjors} that
a specific nonlinear perturbation, corresponding to two insertions of \eqref{leadG}, gives a correction to the potential of the form
\begin{equation} \label{delvfromallnonlin}
\delta V_{\overbar{D3}} \lesssim \mu_3 e^{4 A_{\mathrm{tip}}} \times e^{A_{\mathrm{tip}}}\,,
\end{equation} which can be more important than some of the modes in \eqref{delvfromall}, but less important than the linearized flux perturbation \eqref{leadG}.

Let us summarize. To compute the influence of a gaugino condensate in the ultraviolet on anti-D3-branes in the infrared,
one can allow perturbations of all of the supergravity fields, grading these modes via a spurion analysis, and examine the resulting solution for $\Phi_+$ in the infrared.
We have collected here, in \eqref{delvfromall}, the leading contributions of the fields that appear in \eqref{eqn:phi plus from flux}, at linear order in perturbations.
Results for all fields, to all orders, appear in \cite{Gandhi:2011id,gandhisjors}, and the only nonlinear correction competitive with any of the terms in \eqref{delvfromall} is the quadratic flux perturbation \eqref{delvfromallnonlin}.

The final result is that the largest correction to the anti-D3-brane potential mediated by excitations of the throat solution
is suppressed by at least a factor $e^{\frac{1}{2}A_{\mathrm{tip}}} \ll 1$ compared to the anti-D3-brane potential in the background solution, and so can be neglected.
This finding is compatible with that of \S\ref{d3baroncond}, and constitutes strong evidence for \eqref{thetotTapproxyes}.

\section{Cancellation of Divergences, and the D3-brane Potential}\label{app:d3}

In this appendix we give details of the computation of the four-dimensional curvature $\mathcal{R}_4$.
First, in \S\ref{secc1} we show that the singular terms contributing to the master equation \eqref{eqn:Master Final} cancel each other, and the finite remainder is the scalar potential \eqref{10dans} for the K\"ahler modulus $T$, in exact agreement with the four-dimensional analysis: see \eqref{thecancelgoal}.

Then, in \S\ref{sec:theD3potential} we repeat this calculation for a compactification containing a D3-brane.  In this case the result expected from the four-dimensional theory is the F-term potential \eqref{eqn:vfwithd3} for the D3-brane moduli and the K\"ahler modulus.  We recover this result as well from ten dimensions in \eqref{eqn:potential from flux}.

In summary, the ten-dimensional computations of this appendix yield finite answers for the four-dimensional curvature, in compactifications with or without D3-branes.  These results precisely agree with the corresponding expressions obtained in the associated four-dimensional effective theories.

\subsection{Cancellation of divergences}\label{secc1}

We begin by adapting the master equation \eqref{eqn:Master Final}.  The term in \eqref{eqn:Master Final} involving $\diff_a \Phi_-\diff^a\Phi_-$ is smaller than $\mathcal{O}\bigl(\langle\lambda\lambda\rangle^2\bigr)$, and can be neglected for present purposes.
We likewise omit the kinetic terms for the moduli $u$ and $\tau$.  Following \eqref{totalTmunu}, the trace of the stress-energy tensor $T_{\mu\nu}^{D7}$ of the D7-brane can be written $T_{\mu\nu}^{D7}g^{\mu\nu}=T_{\mu\nu}^{\lambda\lambda}g^{\mu\nu}+T_{\mu\nu}^{\lambda\lambda\lambda\lambda}g^{\mu\nu}\equiv T^{\lambda\lambda}+T^{\lambda\lambda\lambda\lambda}$.
We thus have
\begin{equation}
\begin{aligned}\label{masterfinal2}
M_{\mathrm{pl}}^2\mathcal{R}_4[g]&=-\int_M \sqrt{g_6} e^{-4A}T^{\lambda\lambda}-\int_M \sqrt{g_6} e^{-4A}T^{\lambda\lambda\lambda\lambda}\\
&+16\pi\int_M \sqrt{g_6} e^{-12u+4A}\rho_{\overbar{D3}}-8\pi e^{-8u}\int_M \sqrt{g_6}\mathcal{R}_6[g]\,,\end{aligned}
\end{equation} where we have applied our convention that $(2\pi)^2\alpha'=1$.

Three of the four contributions on the right-hand side of \eqref{masterfinal2} include singular terms.
In the presence of the localized ISD flux \eqref{eqn:flux} sourced by the gaugino condensate, the soft mass term \eqref{eqn:ciugcg} has a singular stress-energy that enters $T_{\mu\nu}^{\lambda\lambda}$.  In the presence of the IASD flux \eqref{eqn:flux from gauginofirst} sourced by the gaugino condensate, the IASD flux kinetic term, proportional to $|G_-|^2$, is likewise singular, and contributes to $\rho_{\overline{D3}}$ via \eqref{rhod3d3bar}.  Finally, the internal curvature $\mathcal{R}_6$ is singular in the presence of singular sources.\footnote{We thank the referee for useful remarks about these contributions.}  Our goal is now to show that \emph{these three singularities cancel}, and the finite remainder is the F-term potential \eqref{10dans}: that is,
\begin{equation}\label{thecancelgoal}
\boxed{\vphantom{\Biggl(\Biggr)}\frac{1}{4}\int_M\sqrt{g_6}\Bigl(-e^{-4A} T^{\lambda\lambda}-e^{-4A}T^{\lambda\lambda\lambda\lambda}+16\pi e^{4A}\rho_{\overline{D3}}-8\pi\mathcal{R}_6[g]\Bigr) = V}\,,
\end{equation} up to corrections smaller than $\mathcal{O}\bigl(\langle\lambda\lambda\rangle^2\bigr)$.

Let us first set our notation.  We will expand in powers of $\langle\lambda\lambda\rangle$, with superscripts $(i)$ denoting quantities of order $\mathcal{O}(\langle\lambda\lambda\rangle^i)$.  We take $G_{(0,3)}$ to be of order $\mathcal{O}(\langle\lambda\lambda\rangle)$.
Capital indices $M,N$ run from $1$ to $6$, while indices $a,b$ run from $1$ to $3$, and we adopt the convention $g^{MN}v_Mv_N=2g^{a\bar{b}}v_a\bar{v}_{\bar{b}}.$
To simplify our expressions, we denote $g^{(1)}_{MN}$ as $h_{MN}$, $g^{(0) MN} g^{(1)}_{MN}$ as $h$, $\mathrm{det}(g^{(0)})$ as $\bar{g}$,
$G_{+ab\bar{c}}^{(0)}$ as $\chi_{ab\bar{c}}$, and $G_{-a\bar{b}\bar{c}}^{(1)}$ by $\eta_{a\bar{b}\bar{c}}$.
We have $\mathcal{R}_6^{(0)}=0$, and we fix the gauge $\partial^M h_{MN}=0.$
The D3-brane and anti-D3-brane charge densities are
\begin{equation}\label{rhod3d3bar}
\rho_{D3}=\frac{1}{2\im\tau}|G_+|^2+\rho_{D3}^{\mathrm{loc}}\,, \qquad
\rho_{\overline{D3}}=\frac{1}{2\im\tau}|G_-|^2+\rho_{\overline{D3}}^{\mathrm{loc}}\,,
\end{equation}
where $\rho_{D3}^{\mathrm{loc}}$ and $\rho_{\overline{D3}}^{\mathrm{loc}}$ are the charge densities due to localized D3-branes and anti-D3-branes, respectively.  For now (in contrast to \S\ref{sec:theD3potential}) we are assuming that there are no localized D3-branes or anti-D3-branes, and so we have $\rho_{D3}=\frac{1}{4\im\tau}\chi_{ab\bar{c}}\bar{\chi}^{ab\bar{c}}$ and $\rho_{\overline{D3}}=\frac{1}{4\im\tau}\eta_{a\bar{b}\bar{c}}\bar{\eta}^{a\bar{b}\bar{c}}.$ In a local coordinate patch, we fix the gauge $\Omega_{abc}=\epsilon_{abc}$ and $\chi_{ab}^{~~c}=\xi\sqrt{\rho_{D3}}\epsilon_{ab}$, for $c \in \{1,2,3\}$, and with $\xi$ a constant.
The equations of motion for this system are well-known, and can be found in, for example, \S3.1 of \cite{Kim:2018vgz}.

Discarding total derivatives and retaining terms up to $\mathcal{O}(\langle\lambda\lambda\rangle^2),$ we have
\begin{align}
\int_M\sqrt{g_6}\,\mathcal{R}_6[g]=-\frac{1}{4}\int_M\sqrt{\bar{g}}\Bigl(\partial_M h \partial^M h-\partial_M h^{NP}\partial^M h_{NP}\Bigr)\,.
\end{align}
The equation of motion for $h_{MN}$ is
\begin{equation}\label{theheom}
\nabla^2 h_{MN}+\nabla_M\nabla_N h=\frac{e^{4A}}{2\im\tau}\Bigl(\chi_{(M}^{~~~PQ}\bar{\eta}_{N)PQ}+c.c.\Bigr),
\end{equation}
where \cite{Baumann:2010sx}
\begin{equation}
\eta_{a\bar{b}\bar{c}}=-i\frac{e^{-4A-\phi/2}\lambda\lambda}{32\pi^2}\partial_a\partial^{\bar{d}}G_{(2)}(z;z_{D7})\overline{\Omega}_{\bar{b}\bar{c}\bar{d}}\,.
\end{equation}
Because $\chi_{ab\bar{c}}$ is a (2,1) form and $\eta_{a\bar{b}\bar{c}}$ is a (1,2) form, \eqref{theheom} implies that $\nabla^2 h=0$.
We will thus take $h=0$, so that \eqref{theheom} takes the form
\begin{equation}
\nabla^2 h_{ab}=\frac{e^{4A}}{\im\tau}\chi_{(a }^{~~\bar{c}d}\bar{\eta}_{b)\bar{c}d}\,, \qquad
\nabla^2 h_{\bar{a}\bar{b}}=\frac{e^{4A}}{\im\tau}\bar{\chi}_{(\bar{a} }^{~~c\bar{d}}\eta_{\bar{b})c\bar{d}}\,.
\end{equation}
Solving in terms of the six-dimensional and two-dimensional Green's functions $G_{(6)}$ and $G_{(2)}$, we find
\begin{align}
h_{ab}=\zeta \int_M d^6x' G_{(6)}(x;x')\chi_{ac\bar{d}}\partial^{c}\partial^{e}G_{(2)}(x';x_{D7})\Omega_{be}^{~~\bar{d}}+(a\leftrightarrow b)\,,
\end{align}
and $h_{\bar{a}\bar{b}}=\overline{h}_{ab}$, where
\begin{equation}
\zeta=-i\frac{e^{\phi/2}}{2^6\pi^2}\,\langle\lambda\lambda\rangle\,.
\end{equation}
We thus find that to $\mathcal{O}(\langle\lambda\lambda\rangle)^2$,
\begin{align}
2\int_M \sqrt{g_6}\mathcal{R}_6[g]=\int_M \sqrt{\bar{g}}\Bigl(-h^{ab}\nabla^2 h_{ab}\Bigr)\,.
\end{align}
We now use an identity that is applicable in the local coordinate chart,
\begin{equation}
\partial_z\partial_{\bar{z}}G_{(2)}(z;0)=\frac{1}{2}g_{z\bar{z}}\left( \delta^{(2)}(z)-\frac{k}{\mathcal{V}_\perp}\right)\,,
\end{equation} where $z$ is the complex coordinate for the space transverse to the D7-brane stack.
We can then simplify $-h^{ab}\nabla^2 h_{ab}$ as follows:
\begin{align}
-h^{ab}\nabla^2 h_{ab}=&-\zeta h^{ab}(x) \left(\chi_{ac\bar{d}}\partial^{c}\partial^{e}G_{(2)}(z;z_{D7})\Omega_{be}^{~~\bar{d}}+(a\leftrightarrow b)\right)\\
=&\frac{|\zeta|^2}{2}\left(\bar{\chi}^{ac\bar{d'}}\partial_{e'}G_{(2)}(z;z_{D7})\overline{\Omega}^{be'}_{~~\bar{d}'}+(a\leftrightarrow b)\right)\left(\chi_{ac\bar{d}}\partial^{e}G_{(2)}(z;z_{D7})\Omega_{be}^{~~\bar{d}}+(a\leftrightarrow b)\right)\label{thesign}\\
=&2^5|\zeta|^2e^{-\phi}\rho_{D3}\partial_e G_{(2)}\partial^eG_{(2)}\,.\label{thecancelgoalr}
\end{align}
To arrive at the sign in \eqref{thesign} we used $\partial_{x'}G_{(6)}(x;x') = -\partial_{x}G_{(6)}(x;x')$.

We next compute $\int_M \sqrt{g_6}e^{4A}\rho_{\overline{D3}}$:
\begin{align}
\int_M \sqrt{\bar{g}}e^{4A}\rho_{\overline{D3}}=&\int_M \sqrt{\bar{g}}\frac{e^{4A}}{\im\tau}\frac{1}{4}\eta_{a\bar{b}\bar{c}}\bar{\eta}^{a\bar{b}\bar{c}}\\
=&|\zeta|^2\int_M \sqrt{\bar{g}} e^{-4A}e^{-\phi}\partial_a\partial^{\bar{d}}G_{(2)}\epsilon_{\bar{b}\bar{c}\bar{d}}\partial^a\partial^dG_{(2)} \epsilon_{bcd}g^{b\bar{b}}g^{c\bar{c}}\\
=&2^4|\zeta|^2\int_M \sqrt{\bar{g}} e^{-4A}e^{-\phi}\partial_a\partial_dG_{(2)}\partial^{a}\partial^dG_{(2)}\\
=&-2^3|\zeta|^2\int_M \sqrt{\bar{g}} \rho_{D3}e^{-\phi}\partial_aG_{(2)}\partial^aG_{(2)},\label{thecancelgoalrho}
\end{align}
where we used $2\partial_a\partial^a e^{-4A}=-\rho_{D3},$ which holds to lowest order.

The final singular contribution comes from the D7-brane action.
From \eqref{tmunu2lambda} and \eqref{llsing}
we have
\begin{equation}
-\frac{1}{4}\int_M \sqrt{g_6} e^{-4A} T^{\lambda\lambda}=\kappa_4^2 \int_X \sqrt{-g_4}e^{\kappa_4^2K}K^{T\overbar{T}}\diff_{\overbar{T}}\overline{W} K_T W+c.c.-S_{\lambda\lambda}^{\mathrm{sing}}\,,
\end{equation} with
\begin{equation}
-S_{\lambda\lambda}^{\mathrm{sing}}=-2\pi \bar{\zeta} \int_M G_{\lambda\lambda}\cdot \Omega\delta^{(0)}+c.c.\,,\label{midsteptll}\\
\end{equation} where $G_{\lambda\lambda}$ is given in \eqref{eqn:flux2}.

To manipulate $S_{\lambda\lambda}^{\mathrm{sing}}$, we derive an identity involving the two-dimensional Green's function.
Taking the internal space transverse to the D7-branes to be compact, with volume $\mathcal{V}_\perp$, Green's equation takes the form
\begin{equation}\label{gfwvolsis}
2g^{a\bar{b}}\diff_a\diff_{\bar{b}}G_{(2)}(z;0)=\delta_{(2)}(z)-\frac{1}{\mathcal{V}_\perp}\,.
\end{equation}
It follows that
\begin{align}
\int_M e^{-4A} \left(\delta^{(0)}-\frac{1}{\mathcal{V}_\perp}\right)^2=&\left(\int_M e^{-4A} \delta^{(0)^2}\right) -\frac{2 e^{4u}\re(T)}{\mathcal{V}_\perp}+\frac{\mathcal{V}}{\mathcal{V}_\perp^2}\\
=&\left( \int_M e^{-4A}\delta^{(0)^2}\right)-\frac{e^{4u}\re(T)}{\mathcal{V}_\perp}\\
=&\int_M e^{-4A}\left(\delta^{(0)}-\frac{1}{\mathcal{V}_\perp}\right)\delta^{(0)},
\end{align}
which implies that
 \begin{equation}\label{del2id}
\int_M e^{-4A}\partial_a\partial^a G_{(2)}(z;z_{D7}) \delta^{(0)}=\int_M 2e^{-4A}\partial_a\partial^a G_{(2)}(z;z_{D7})\partial_b\partial^b G_{(2)}(z;z_{D7}).
\end{equation}
Using \eqref{eqn:flux2} in \eqref{ssingis} and using \eqref{del2id}, we find
\begin{align}
-S_{\lambda\lambda}^{\mathrm{sing}}=&-\int_M 2^7\pi|\zeta|^2e^{-4A}e^{-\phi}\partial_a\partial^aG_{(2)}(z;z_{D7})\partial_b\partial^bG_{(2)}(z;z_{D7})\label{finaltll}\\
=&\int_M 2^6 \pi |\zeta|^2 e^{-\phi} \rho_{D3}\partial_aG_{(2)}(z;z_{D7})\partial^aG_{(2)}(z;z_{D7}).\label{reallyfinalt11}
\end{align}

Combining \eqref{thecancelgoalr}, \eqref{thecancelgoalrho}, and \eqref{reallyfinalt11}, we find that
\begin{equation}\label{thecancelgoalpart}
\frac{1}{4}\int_M\sqrt{g_6}\Bigl(-e^{-4A} T^{\lambda\lambda}+16\pi e^{4A}\rho_{\overline{D3}}-8\pi\mathcal{R}_6[g]\Bigr) = V_{\lambda\lambda}\,,
\end{equation} where the finite term $V_{\lambda\lambda}$ was given in \eqref{twogauginoV}.  Including also the finite term resulting from $T_{\mu\nu}^{\lambda\lambda\lambda\lambda}$, see \eqref{eqn:fourfermiVterm},
we arrive at \eqref{thecancelgoal}, completing the proof.

\subsection{D3-brane potential from flux}\label{sec:theD3potential}

We now turn to the case in which a spacetime-filling D3-brane is present.  The potential for motion of a D3-brane in a nonperturbatively-stabilized flux compactification, such as \cite{KKLT},
is well understood from the perspective of the four-dimensional effective supergravity theory \cite{Kachru:2003sx,Baumann:2006th,Krause:2007jk,Baumann:2007ah}, with the K\"ahler potential obtained in \cite{DeWolfe:2002nn} (see also \cite{Martucci:2014ska,Cownden:2016hpf,Martucci:2016pzt}) and with the nonperturbative superpotential computed in \cite{Berg:2004ek,Baumann:2006th}.  Showing that this potential is reproduced by the Dirac-Born-Infeld + Chern-Simons action of a probe D3-brane in a candidate
ten-dimensional solution sourced by gaugino condensation serves as a quantitative check of the ten-dimensional configuration \cite{Koerber:2007xk,Baumann:2010sx,Dymarsky:2010mf}.
An exact match was demonstrated in \cite{Baumann:2010sx} in the limit that four-dimensional gravity decouples.

In this appendix we compute the potential of such a D3-brane probe.
Through a consistent treatment of the Green's functions on the compact space, we extend the match found in \cite{Baumann:2010sx} to include terms proportional to $\kappa_4^2$.

Within this appendix we take the K\"ahler potential \eqref{kpotis} to include D3-brane moduli,
\begin{equation}\label{kdwgpotis}
\kappa_4^2K=-3\log\bigl(T+\overbar{T}-\gamma k\bigr)-\log\bigl(-i(\tau-\overbar{\tau})\bigr)-\log\left(i\int_M e^{-4A}\Omega\wedge\overline{\Omega} \right)+\log\Bigl(2^7 \mathcal{V}^3\Bigr)\,,
\end{equation}
with (cf.~\cite{DeWolfe:2002nn,Baumann:2007ah,Cownden:2016hpf})\footnote{As explained in \cite{Baumann:2007ah}, the relation \eqref{defofgamma} should be understood to hold exactly at a reference location in field space.  Deviations from \eqref{defofgamma} at other locations lead to
corrections of order $\frac{\gamma k}{T+\overbar{T}}$ in \eqref{phiminflux2pt5} and \eqref{phiminflux3} below, which we will neglect.}
\begin{equation}\label{defofgamma}
\gamma=\frac{2}{3}\mu_3 \kappa_4^2\, \text{Re}(T)e^{-4u}=\frac{1}{3\mathcal{V}_\perp}\,.
\end{equation}
Here $k$ is the K\"ahler potential of $M$, obeying $k_{a\bar{b}}=g_{a\bar{b}}$,
where $a$ and $\bar{b}$ are holomorphic and anti-holomorphic indices for D3-brane moduli.  We use the convention $ds^2=2g_{a\bar{b}} dz^a d\bar{z}^{\bar{b}}$ for the line element.

The $G_-$ flux sourced by gaugino condensation \cite{Baumann:2010sx} is given by \eqref{eqn:flux from gauginofirst},
where $G_{(2)}$ is the Green's function on the internal space transverse to the D7-branes.
If this space is taken to be noncompact, we have
\begin{equation}
G_{(2)}(z;0)=\frac{1}{2\pi}\log|z|\,,
\end{equation} in terms of a local coordinate $z$.

The flux \eqref{eqn:flux from gauginofirst} is a source for the scalar $\Phi_-$, leading to a potential for D3-brane motion.
The equation of motion for $\Phi_-$ is
\begin{equation}
\nabla^2\Phi_-=\frac{e^{8A}}{\im\tau}|G_-|^2+\ldots \label{eqn:phi minus from flux}
\end{equation} where the omitted terms are not important for the present computation.
Solving  \eqref{eqn:phi minus from flux} and taking the D7-brane location to be given by an equation $h(z)=0$ in local coordinates,
one finds\footnote{Throughout this appendix, we write only the contribution to $\Phi_-$ sourced by $G_-$ flux via \eqref{eqn:phi minus from flux}.  Further contributions are present in general \cite{Baumann:2010sx}.}
\begin{align} \label{rigidsol}
\Phi_-=&\int_M d^6y\, G_{(6)}(z;z') \frac{e^{8A}}{\im\tau}|G_-|^2\\
=&\frac{e^{\kappa_4^2 K}e^{16u}}{4\pi^2N_c^2} g^{a\bar{b}} \frac{\diff_a h\diff_{\bar{b}}\bar{h}}{h\bar{h}}|W_{\mathrm{np}}|^2\,,
\end{align}
so that
\begin{equation}\label{therig}
\mu_3 \Phi_-= e^{12u}  e^{\kappa_4^2K} K^{a\bar{b}} \diff_a W\diff_{\bar{b}}\overline{W}\,.
\end{equation}
Thus, the flux \eqref{eqn:flux from gauginofirst} sourced by gaugino condensation gives rise to a profile for $\Phi_-$ that matches the \emph{rigid part} of the F-term potential.

At this point, the K\"ahler connection terms in the F-term potential are not evident in the ten-dimensional computation.
The result of this appendix, which we will now establish, is that the K\"ahler connection terms arise once one consistently incorporates finite volume effects in the Green's function.

If the space transverse to the D7-branes is compact, with volume $\mathcal{V}_\perp$, then the Green's function reads
\begin{equation}\label{thetrueG}
G_{(2)}(z;0)=\frac{1}{2\pi}\log|z|-\frac{k}{6\mathcal{V}_\perp}\,.
\end{equation}
Using \eqref{thetrueG} to solve \eqref{eqn:phi minus from flux},
one finds
\begin{align}
\Phi_{-}=& \int_M G_{(6)}(z;z')\diff_a \diff_b G_{(2)}(z';z_{D7})\diff_{\bar{a}}\diff_{\bar{b}}G_{(2)}(z';z_{D7})g^{a\bar{a}}g^{b\bar{b}} e^{16u}\left|\frac{\lambda\lambda}{32\pi^2}\right|^2 \Omega\cdot\overline{\Omega}\\
=&\frac{1}{2} \int_M  \delta_{(6)}(z;z')\diff_a G_{(2)}(z';z_{D7})\diff_{\bar{b}}G_{(2)}(z';z_{D7})g^{a\bar{b}}e^{16u} \left|\frac{\lambda\lambda}{32\pi^2}\right|^2 \Omega\cdot\overline{\Omega}\\
=&2^4 |\zeta|^2e^{-\phi}\partial_aG_{(2)}(z;z_{D7})\partial^aG_{(2)}(z;z_{D7})\label{phiminfromfl3}\\
=&\frac{1}{4N_c^2 \pi^2} \left( \frac{\diff_a h(z)}{h(z)}-\frac{2\pi k_a}{3\mathcal{V}_\perp}\right)\left( \frac{\diff_{\bar{b}} \bar{h}(\bar{z})}{\bar{h}(\bar{z})}-\frac{2\pi k_{\bar{b}}}{3\mathcal{V}_\perp}\right)g^{a\bar{b}}e^{\kappa_4^2 K}e^{16u}|W_{\mathrm{np}}|^2\,.\label{phiminfromfl}
\end{align}
The F-term potential that we wish to compare to \eqref{phiminfromfl} is given by
\begin{equation}
V_F = e^{\kappa_4^2K}\Bigl(K^{\Delta\overline{\Gamma}}D_\Delta WD_{\overline{\Gamma}}\overline{W}-3\kappa_4^2 W\overline{W}\Bigr)\,,\label{eqn:vfwithd3}
\end{equation} where $K^{\Delta \overline{\Gamma}}$ is the inverse K\"ahler metric derived from the DeWolfe-Giddings K\"ahler potential \cite{DeWolfe:2002nn,Burgess:2006cb},
\begin{equation}
K^{\Delta \overline{\Gamma}}=\frac{\kappa_4^2 (T+\overbar{T}-\gamma k)}{3\gamma}\left(\begin{array}{c|c}\gamma (T+\overbar{T}-\gamma k)+ \gamma^2 k_a k^{a\bar{b}}k_{\bar{b}}& \gamma k_a k^{a\bar{b}}\\ \hline \gamma k^{a\bar{b}}k_{\bar{b}} & k^{a\bar{b}}\end{array}\right),
\end{equation}
and the index $\Delta$ runs over $T$ and the D3-brane moduli $y_a$.
Using \eqref{defofgamma}, we can rewrite \eqref{phiminfromfl} as
\begin{align} \label{phiminflux2pt5}
\Phi_{-}=& \frac{e^{\kappa_4^2K}e^{16u}}{4\pi^2}g^{a\bar{b}}\left(D_a W+ \gamma k_aD_TW\right)\left(D_{\bar{b}} \overline{W}+\gamma k_{\bar{b}}D_{\overbar{T}}\overline{W} \right)+\ldots\\
=&e^{\kappa_4^2K}e^{12u} \frac{\kappa_4^2\text{Re}(T)}{3\pi\gamma} \left(\begin{array}{c c} D_T W & D_a W\end{array}\right)\left(\begin{array}{c|c} \gamma^2 k_a k^{a\bar{b}}k_{\bar{b}}& \gamma k_a k^{a\bar{b}}\\ \hline \gamma k^{a\bar{b}}k_{\bar{b}} & k^{a\bar{b}}\end{array}\right) \left(\begin{array}{c}D_{\overbar{T}}\overline{W}\\ D_{\bar{b}}\overline{W}\end{array} \right)+\ldots\,, \label{phiminflux3}
\end{align}
where the omitted terms are of higher order in $\frac{\gamma k}{T+\overbar{T}}.$

Combining \eqref{10dans} and \eqref{phiminflux3}, we conclude that in a compact space, the flux \eqref{eqn:flux from gauginofirst} sourced by gaugino condensation leads to a $\Phi_-$ profile that agrees with the F-term potential \eqref{eqn:vfwithd3}:
\begin{equation}
\mu_3e^{-12u}\Phi_-(z)+V_{\lambda\lambda}+V_{\lambda\lambda\lambda\lambda}=e^{\kappa_4^2K}\Bigl(K^{\Delta\overline{\Gamma}}D_\Delta WD_{\overline{\Gamma}}\overline{W}-3\kappa_4^2 W\overline{W}\Bigr)+\ldots\,,\label{eqn:potential from flux}
\end{equation} where again the omitted terms are subleading in $\frac{\gamma k}{T+\overbar{T}}.$

Finally, we will show that \eqref{eqn:potential from flux} also follows from \eqref{eqn:Master Final} upon adapting the calculation of \S\ref{secc1} to account for the presence of localized D3-branes.
From \eqref{theheom} we see that the metric at order $\mathcal{O}(\langle\lambda\lambda\rangle)$ is only sourced by the fluxes $\chi$ and $\eta,$ and so \eqref{thecancelgoalr} is altered to
\begin{equation}
2\int_M\sqrt{g_6}\mathcal{R}_6[g]=\int_M2^5|\zeta|^2e^{-\phi}\bigl(\rho_{D3}-\rho_{D3}^{\mathrm{loc}}\bigr)\partial_e G_{(2)}\partial^eG_{(2)}\,.\label{c3r6}
\end{equation}
On the other hand, the warp factor $e^{-4A}$ is sourced by the full D3-brane charge density $\rho_{D3}$, i.e.~by both localized and distributed sources, and obeys $2\partial_a\partial^a e^{-4A}=-\rho_{D3}$.
As a result, equations \eqref{thecancelgoalrho} and \eqref{reallyfinalt11} continue to hold.

Combining \eqref{c3r6}, \eqref{thecancelgoalrho}, and \eqref{reallyfinalt11} we find
\begin{equation}
\frac{1}{4}\int_M\sqrt{g_6}\Bigl(-e^{-4A} T^{\lambda\lambda}+16\pi e^{4A}\rho_{\overline{D3}}-8\pi\mathcal{R}_6[g]\Bigr) = V_{\lambda\lambda}+2\pi\int_M\sqrt{g_6}\rho_{D3}^{\mathrm{loc}}\Phi_-\,.\label{semifinal c3}
\end{equation}
When $-\frac{1}{4}\int_M \sqrt{g_6}e^{-4A}T^{\lambda\lambda\lambda\lambda}$ is added to \eqref{semifinal c3}, we recover the full F-term potential \eqref{eqn:potential from flux}.

\end{appendix}
\bibliographystyle{JHEP}
\bibliography{refs}
\end{document}